\documentclass[final]{amsart}
\usepackage{fullpage}
\usepackage{graphicx} 
\usepackage{pgfplots}
\pgfplotsset{width=7cm,compat=1.14}
\usepackage{pgfplotstable}
\usepackage{pgffor} 
\pgfplotsset{compat=newest}
\usepgfplotslibrary{groupplots} 
\usepackage{hyperref}
\usepackage{xcolor} 
\usepackage{amsmath} 
\usepackage{geometry} 
\usepackage{comment}
\usepackage{tikz}
\usetikzlibrary{intersections}
\usetikzlibrary{arrows,snakes,shapes}
\usepgfplotslibrary{fillbetween}
\usepackage{caption}
\usepackage{subcaption}
\usepackage{tristan}
\usepackage{booktabs}

\usepackage{bbold}    
\usepackage{csquotes} 
\usepackage[sorting=none,maxnames=1,giveninits=true,eprint=false,url=false]{biblatex} 
\title{Efficient Proton Transport Modelling for Proton Beam Therapy
  and Biological Quantification}

\author{Ben S. Ashby$^{1}$}
\author{Veronika Chronholm$^{2}$}
\author{Daniel K. Hajnal$^{2}$}
\author{Alex Lukyanov$^{3}$}
\author{Katherine MacKenzie$^{1}$}
\author{Aaron Pim$^{2}$}
\author{Tristan Pryer$^{1,2}$}

\address{$^1$Institute of Mathematical Innovation, University of Bath}
\address{$^2$Department of Mathematical Sciences, University of Bath}
\address{$^3$Department of Mathematics and Statistics, University of Reading}

\begin{document}

\begin{abstract}
  In this work, we present a fundamental mathematical model for proton
  transport, tailored to capture the key physical processes
  underpinning Proton Beam Therapy (PBT). The model provides a robust
  and computationally efficient framework for exploring various
  aspects of PBT, including dose delivery, linear energy transfer,
  treatment planning and the evaluation of relative biological
  effectiveness. Our findings highlight the potential of this model as
  a complementary tool to more complex and computationally intensive
  simulation techniques currently used in clinical practice.
\end{abstract}

\maketitle

\section{Introduction}

\begin{figure}[h!]
\includegraphics[width=0.75\linewidth]{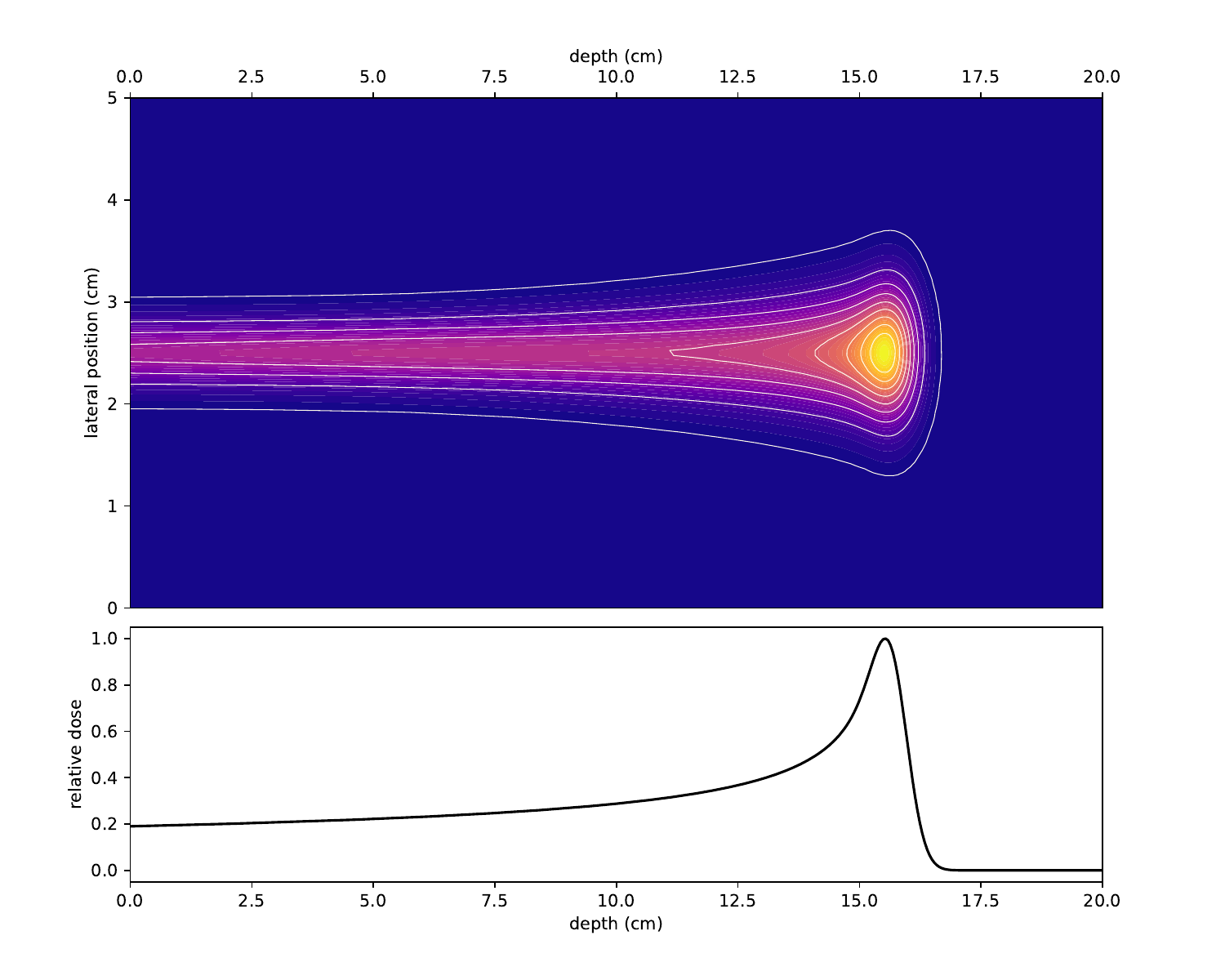}
\caption{
\label{fig:braggpeak}
Simulated dose profile of a proton beam illustrating the Bragg
peak. The simulation was performed using MCsquare with $1.21 \cdot
10^7$ particles, a beam width of 2 mm, without nuclear
interactions. The initial proton energy is 150 MeV with a 1\% energy
spread, and the dose is integrated along the plane orthogonal to the
beam axis.}
\end{figure}

Proton Beam Therapy (PBT) has emerged as an important modality in the
treatment of specific challenging cancers, particularly where
conventional photon-based radiotherapy struggles to minimise
irradiation to surrounding critical tissues. Pediatric cancers, skull
base tumours, and complex head and neck malignancies are example cases
where PBT offers a distinct advantage due to its ability to deposit
energy with precision, peaking at the Bragg peak, see Figure
\ref{fig:braggpeak}.

The Bragg peak enables precise targeting of tumours while sparing
adjacent healthy tissue, making PBT an attractive option for complex
clinical scenarios. However, despite its potential for superior dose
profiles, there are fundamental challenges in treatment planning,
verification, and uncertainty quantification. A key issue is the
variability in patient anatomy during the treatment course, such as
changes in water retention or tumour motion, which can lead to
suboptimal tumour coverage or increased irradiation of healthy
tissues. These challenges limit the potential of PBT to achieve
superior therapeutic outcomes consistently.

Moreover, the intricacies of Relative Biological Effectiveness (RBE),
which varies with proton energy and penetration depth, add a layer of
complexity to treatment optimisation. The RBE, a measure of the
relative damage caused by protons compared to photons, depends heavily
on Linear Energy Transfer (LET), the rate at which energy is deposited
along the proton track. Regions with higher LET are known to cause
more complex and clustered DNA damage, which is harder for cells to
repair, leading to increased biological effectiveness. This
correlation makes LET a key metric for linking physical dose
distributions to biological outcomes. As highlighted by Nystrom et
al. \cite{nystrom2020treatment}:
\begin{quote}
  ``We believe that the endless discussions of which is the most
  appropriate RBE-model and the exact values of the RBE for different
  tissue and in different parts of the dose distribution should be put
  on hold. Rather, efforts should be put in the development of
  clinically useful tools to visualise LET distributions and the
  possibility to include LET in the optimisation of proton treatment
  plans."
\end{quote}

This sentiment underscores the urgent need for practical,
clinically-relevant methodologies to optimise PBT treatment plans
while navigating biological complexities. In this work, we adopt a
fundamental mathematical perspective to address some of these
challenges in PBT treatment planning.  Specifically, we focus on
developing a robust and accessible framework for optimising dose
distribution, incorporating considerations of RBE and its impact on
tumour cell survival. A particular emphasis is placed on LET, enabling
the integration of biological metrics into the optimisation process.

Our approach introduces:
\begin{itemize}
    \item A simplified, yet precise, model that facilitates rapid
      exploration of treatment plan precision and its implications for
      therapeutic outcomes.
    \item A rigorous, accessible, mathematical formulation for
      optimising the biological effective dose, explicitly accounting
      for LET distributions and RBE variability.
    \item A biologically informed framework for interpreting treatment
      plans, incorporating cell survival fraction and RBE as key
      metrics.
    \item A detailed sensitivity analysis of the proposed framework,
      identifying the main parameters that influence treatment
      outcomes.
\end{itemize}
The goal is to provide tools that are not only theoretically sound but
also practical for clinical implementation, bridging the gap between
mathematical modelling and real-world PBT optimisation. This framework
supports investigations towards personalised and adaptive therapy strategies, with the end goal of enhancing
precision in challenging clinical scenarios by enabling the
integration of spatially resolved biological metrics, such as LET,
into treatment planning.

Modelling proton transport and dose distribution in PBT has
traditionally relied on Monte Carlo simulations
\cite{salvat2013generic,jabbari2014fast}, which provide detailed
physical insights but come with significant computational costs
\cite{unkelbach2016reoptimization}. These methods remain the gold
standard for accuracy but are impractical for real-time treatment
planning. Recent advancements in neural network-based methodologies
show promise for accelerating dose prediction by leveraging large
datasets \cite{frizzelle2022using, fanou2023implementation}. However,
these approaches are still nascent in clinical contexts and require
extensive validation.

From a biological perspective, the role of RBE in treatment
optimisation has been an important point of research. Unlike photon
radiotherapy, where energy independence simplifies the calculation of
biologically effective dose (BED) \cite{bellamy2015empirical}, proton
radiotherapy demands careful calibration to account for RBE
variability with depth and energy \cite{chaudhary2014relative}. While
multiple RBE models exist, the lack of consensus on their clinical
applicability continues to pose challenges, and in practice virtually
all treatment centres use a constant RBE of 1.1
\cite{gerweck1999relative, underwood2016variable,
  giantsoudi2013linear, paganetti2019report}, despite the fact that
there is broad consensus that RBE varies along the particle track,
increasing near the Bragg peak \cite{hojo2017difference} to values
significantly larger than 1.1. Indeed, RBE typically reaches values of
approximately 1.6 in the falloff region of the Bragg peak
\cite{paganetti2014relative}.

Finally, efforts to incorporate LET distributions into treatment
planning are gaining traction (see \cite{unkelbach2016reoptimization}
for an example). These efforts align with the broader push toward
integrating data-driven methodologies, such as neural networks, which
have shown promise in accelerating dose prediction and enhancing
planning workflows. Our framework complements such approaches by
offering a rigorous and interpretable model for evaluating and
validating biologically informed treatment plans, as well as providing
tools to visualise the effect that varying optimisation routines to
include additional quantities such as LET has upon the final plan.

It is worth noting that treatment planning is only one aspect of the
broader radiotherapy workflow, which encompasses tumour growth
modelling, temporal variations and interactions with other treatment
modalities such as chemotherapy. While these dynamic aspects can be
addressed using optimal control formulations
\cite{schattler2015optimal}, this work focuses on stationary problems,
providing a foundational framework that enables the capture of key
physical and biological processes. In clinical practice, temporal
changes such as tumour shrinkage or movement are addressed through
adaptive radiotherapy, which relies on re-imaging, re-planning, or
pre-optimised scenarios to ensure treatment quality across long,
fractionated courses of therapy \cite{sonke2019adaptive}.

In this context, the computational efficiency of our framework offers
a significant advantage, particularly in adaptive workflows where
rapid dose calculations and fast treatment plan optimisations are
important. In this work we aim to bridge the gap between theoretical
modelling and clinical applicability, laying the groundwork for
examining biological metrics in current planning workflows.

The rest of this paper is structured as follows: in \S\ref{sec:model},
we introduce the fundamental mathematical model, describing the
principles of charged particle transport, stopping power, and LET and
validate the model against existing Monte Carlo codes. In
\S\ref{sec:biological}, we shift focus to biological metrics,
exploring how the model can incorporate key measures such as the cell
survival fraction and RBE to evaluate the biological effectiveness of
proton beams. \S\ref{sec:uncertainties} addresses model uncertainties,
providing a sensitivity analysis to quantify how variations in key
parameters affect dose and LET predictions. Finally,
\S\ref{sec:optimisation} explores applications to treatment planning,
demonstrating how the framework can optimise dose delivery and
integrate biological metrics to enhance therapeutic outcomes.

\section{Modelling of charged particle transport}
\label{sec:model}

In this section we introduce fundamental modelling concepts in proton
transport and discuss a fundamental model to aid in the exploration of
the ideas discussed here.

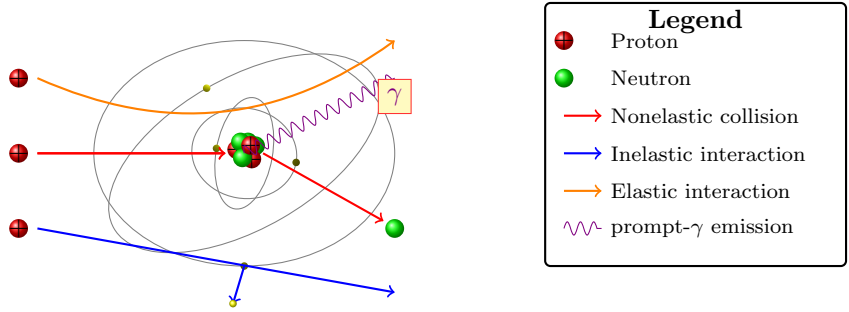
\begin{figure}[h!]
    \centering
    \begin{tikzpicture}[scale=0.5]
      \nucleus \electron{1.5}{0.75}{80} \electron{1.2}{1.4}{260}
      \electron{4}{2}{30} \electron{4}{3}{180}
      \protoncollision{-6.}{0.}{160} \inelastic{-6.}{-2.}
      \elastic{-6.}{2.}

      \begin{scope}[shift={(8,-1)}] 
        \draw [thick,rounded corners=2pt] (0,-2) rectangle (8,5); 
        \node at (4, 4.5) {\textbf{Legend}};        
        \proton{0.5, 4}
        \node[anchor=west, font=\footnotesize] at (1.5,4) {Proton}; 
        \neutron{0.5, 3}
        \node[anchor=west, font=\footnotesize] at (1.5,3) {Neutron};         
        \draw[->,thick,red] (0.5,2) -- +(1,0);
        \node[anchor=west, font=\footnotesize] at (1.5,2) {Nonelastic collision};        
        \draw[->,thick,blue] (0.5,1) -- +(1,0);
        \node[anchor=west, font=\footnotesize] at (1.5,1) {Inelastic interaction};         
        \draw[->,thick,orange] (0.5,0) -- +(1,0);
        \node[anchor=west, font=\footnotesize] at (1.5,0) {Elastic interaction}; 
        \draw[snake=coil, line after snake=0pt, segment aspect=0,
          segment length=5pt,color=red!50!blue] (0.5,-1) -- +(1,0);
        \node[anchor=west, font=\footnotesize] at (1.5,-1) {prompt-$\gamma$ emission};
      \end{scope}
\end{tikzpicture}
    \caption{\em The three main interactions of a proton with matter. A
      \textcolor{red}{nonelastic} proton-nucleus collision, an
      \textcolor{blue}{inelastic} Coulomb interaction with atomic
      electrons and \textcolor{orange}{elastic} Coulomb scattering with
      the nucleus.
      \label{fig:atom}
    }
    \label{fig:diagram_of_scattering}
\end{figure}

\subsection{A model for proton transport}

In this section, we introduce a simplified model for proton transport
that builds upon the fundamental principles shown in Figure
\ref{fig:diagram_of_scattering}. To that end, consider a bounded
domain $X \subset \mathbb{R}^3$. For $0 < E_{\min} < E_{\max}$, we
define the interval $I = [E_{\min}, E_{\max}]$ as the set of
admissible particle energies and let $\mathbb{S}^2$ represent the unit
sphere, which describes possible particle trajectories.

At any given position $\vec{x} \in X$, with energy $E \in I$ and a
trajectory direction $\vec{\omega} \in \mathbb{S}^2$, we are
interested in modelling the particle fluence. The fluence, denoted
$\psi(\vec{x}, E, \vec{\omega})$, describes the differential number of
particles passing through a small surface area within an infinitesimal
energy range. We let $\mathscr S$ denote the \emph{stopping power},
formally defined in \S \ref{sec:stopping}, as a function of the
particle energy in a homogeneous domain. To simplify the analysis, we
assume that angular scattering of the particles is minimal, that
nonelastic collisions are rare. Based on these assumptions, we arrive
at a transport model where the fluence satisfies
\begin{equation}
 \vec{\omega} \cdot \nabla_{\vec{x}} \psi(\vec{x}, E, \vec{\omega})
  +
  \dfrac{\partial}{\partial E}
  \qp{\mathscr S(\vec{x}, E) \psi(\vec{x}, E, \vec{\omega})}
  =
  0.
\end{equation}
This equation captures the balance between the particle's motion
through the medium and the energy loss through ionisation described by
the stopping power.

Boundary conditions are required to close the model. The inflow
condition specifies the fluence at the boundary of the domain
\begin{equation}
  \partial X_- := \ensemble{\vec x\in \partial X}{\vec{n}(\vec{x}) \cdot \vec{\omega} < 0}
\end{equation}
where particles enter, while the energy cutoff condition ensures that
no particles exist above the maximum energy, these read
  \begin{equation}
    \begin{split}
      \psi(\vec{x}, E, \vec{\omega})
      &=
      \mathscr G(\vec{x}, E, \vec{\omega}) \Foreach \vec{x} \in \partial X_-, E\in I
      \\
      \psi(\vec{x}, E_{\max}, \vec{\omega}) &= 0 \Foreach \vec{x} \in X,  \vec{\omega} \in \mathbb{S}^2.
    \end{split}
  \end{equation}
  This formulation provides the foundation we use to explore proton
  transport.

\subsection{Model simplification.}

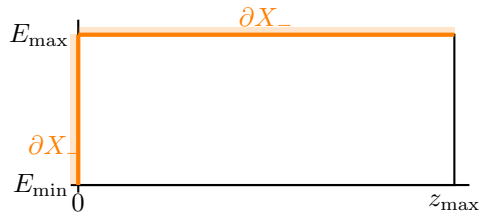
\begin{figure}[h!]
  \begin{center}
    \begin{tikzpicture}
    \def\zmax{5}
    \def\Emin{0.5}
    \def\Emax{2.5}

    \draw[thick] (0, \Emin) rectangle (\zmax, \Emax);

    \draw[thick] (-0.1, \Emin) -- (\zmax + 0.2, \Emin);
    \node[below] at (0, \Emin) {0};
    \node[below] at (\zmax, \Emin) {$z_{\text{max}}$};

    \draw[thick] (0, \Emin - 0.1) -- (0, \Emax + 0.2);
    \node[left] at (0, \Emin) {$E_{\text{min}}$};
    \node[left] at (0, \Emax) {$E_{\text{max}}$};

    \draw[orange, ultra thick] (0, \Emin) -- (0, \Emax);  
    \draw[orange, ultra thick] (0, \Emax) -- (\zmax, \Emax); 

    \node[orange] at (-0.3, 1) {$\partial X_-$}; 
    \node[orange] at (\zmax/2, \Emax + 0.2) {$\partial X_-$}; 

    \fill[orange, opacity=0.2] (0, \Emin) rectangle (-0.1, \Emax); 
    \fill[orange, opacity=0.2] (0, \Emax) rectangle (\zmax, \Emax + 0.1); 

\end{tikzpicture}    
    \caption{\label{fig:domain} An illustration of the domain and
      the relevant inflow boundary.}
  \end{center}
\end{figure}

Given the assumptions made, it is useful to consider the variable
\begin{equation}
    z := (\vec x - \vec x_0)\cdot \vec \omega, 
\end{equation}
where $\vec x_0 \in \partial X_-$ is a given entrance point and $\vec
\omega \in \mathbb{S}^2$ is a given trajectory. We define
\begin{equation}
  z_{\max} := \inf\qc{s >0: ~ \vec x_0 + s \vec \omega \in \partial X, ~ \vec n(\vec x_0 + s \vec \omega ) \cdot \vec \omega > 0},
\end{equation}
which, for a given entrance point, represents the minimal path length
to an outflow boundary. Then, for fixed $\vec x_0$ and $\vec \omega$,
we define the one-dimensional stopping power and fluence in terms of
the energy $E$ and the one-dimensional track length $z$ by the
following change of variables,
\begin{equation}
    \begin{split}
        S(z,E) &:= \mathscr S(\vec x_0 + z \vec \omega, E)  
        \\
        u(z,E)
        &:=
        \psi(\vec x_0 + z \vec \omega, E, \vec \omega)
        \\
        g(E) &:= \mathscr G(\vec x_0, E, \vec \omega) .
    \end{split}
\end{equation}
We obtain the following problem for $u$:
\begin{equation}\label{eq:1dpde}
  \dfrac{\partial }{\partial z}u(z, E)
  +
  \dfrac{\partial }{\partial E}\qp{S(z, E)u(z, E)} =0, \qquad \forall z,E \in (0, z_{\max})\times I,
\end{equation}
subject to the boundary conditions
\begin{equation}
  \begin{split}
    u(0, E) &= g(E),~ \forall E \in I, \qquad
    \\
    u(z,E_{\max}) &= 0, ~ \forall z \in (0, z_{\max}).
  \end{split}
\end{equation}
Figure \ref{fig:domain} gives a visualisation of the domain and inflow boundaries.

\subsection{Energy, range \& stopping power}
\label{sec:stopping}

Suppose that a proton beam consisting of particles of a single energy
$E_0$ enter a medium.  There is a fundamental relationship between
$E_0$ (in MeV) and the range $R_0$ the protons penetrate into the
medium. This relationship is often modelled as a power law
\begin{equation}\label{eq:range_energy}
    R_0 = \alpha E_0^p,
\end{equation}
where $p \in [1,2]$ and $\alpha > 0$ is related to the mass density of
the medium. Some indicative empirical values are given in Table
\ref{tab:range_energy}.
\begin{table}[h!]
  \centering
  \begin{tabular}{lrr}
    \toprule
    Medium &    $p$ &  $\alpha$ \\
    \midrule
    Water & 1.75$\pm$0.02 & 0.00246$\pm$0.00025 \\
    Muscle & 1.75 &            0.0021 \\
    Bone & 1.77 &            0.0011 \\
    Lung & 1.74 &            0.0033 \\
    \bottomrule
  \end{tabular}
  \caption{
    \label{tab:range_energy}
    Range-Energy relationship parameters for different media. Notice
    the parameter $p$ remains relatively constant across different
    biological media. In contrast, the parameter $\alpha$ varies more
    significantly, as it is strongly dependent on the density and
    composition of each medium. The uncertainty in the water phantom
    is based on comparing three parameterisations of the Bragg Kleeman
    rule from \cite{pettersen2018accuracy, boon1998dosimetry,
      bortfeld1997analytical}.  }
\end{table}

From this relationship, one can derive a formula for the remaining
energy $E(z)$ at a given depth, $z \geq 0$ by observing that at depth
$z$, the range of the beam is $R_0 - z$. Applying the range-energy
relationship \eqref{eq:range_energy} at this depth yields
\begin{equation}
    R_0 - z = \alpha E(z)^p,
\end{equation}
Solving for $E$ we see
\begin{equation}\label{eq:E_of_x}
  E(z) 
  =
  \alpha^{-\frac 1 p}\left( R_0 - z\right)^{\frac 1 p}.
\end{equation}
This expression describes the energy of the proton beam as a function
of depth. The linear stopping power, defined as the energy loss per
unit distance travelled, can then be computed by
\begin{equation}
  S(z)
  :=
  -\frac{\d E(z)}{ \d x}
  =
  \frac{\alpha^{-\frac{1}{p}}}{p}\left( R_0 - z\right)^{\frac 1 p - 1}.
\end{equation}
Finally, since the relationship \eqref{eq:E_of_x} is invertible for $0
\leq z \leq R_0$, the stopping power can be expressed in terms of
energy:
\begin{equation}
  \label{eq:BraggKleeman}
  S(E) = \frac{1}{\alpha p}E^{1-p}.
\end{equation}
This representation, called the Bragg-Kleeman rule, illustrates how
the stopping power decreases as a function of energy which is the
property that yields the forward facing peaked nature of the Bragg
peak.

\begin{remark}[Relativistic effects]
  Proton therapy typically uses proton energies in the range of $50$
  to $250$ MeV. While these energies are high enough to require
  accurate modeling of stopping power, they are not so high that
  relativistic effects dominate. In this intermediate energy range,
  the Bragg-Kleeman model provides a sufficiently accurate
  approximation of stopping power while maintaining simplicity and
  computational efficiency \cite{ulmer2007theoretical}.

  The Bethe-Bloch formula, illustrated in Figure \ref{fig:bethebloch},
  is more precise at relativistic speeds as it incorporates
  corrections and additional parameters necessary at very high
  energies \cite{navas2024review}. However, it also introduces
  complexity to calculations and exhibits unphysical behaviour at low
  energy levels, making it less suitable for the energy ranges used in
  proton therapy. 
  \begin{figure}[h!]
    \includegraphics[width=0.5\textwidth]{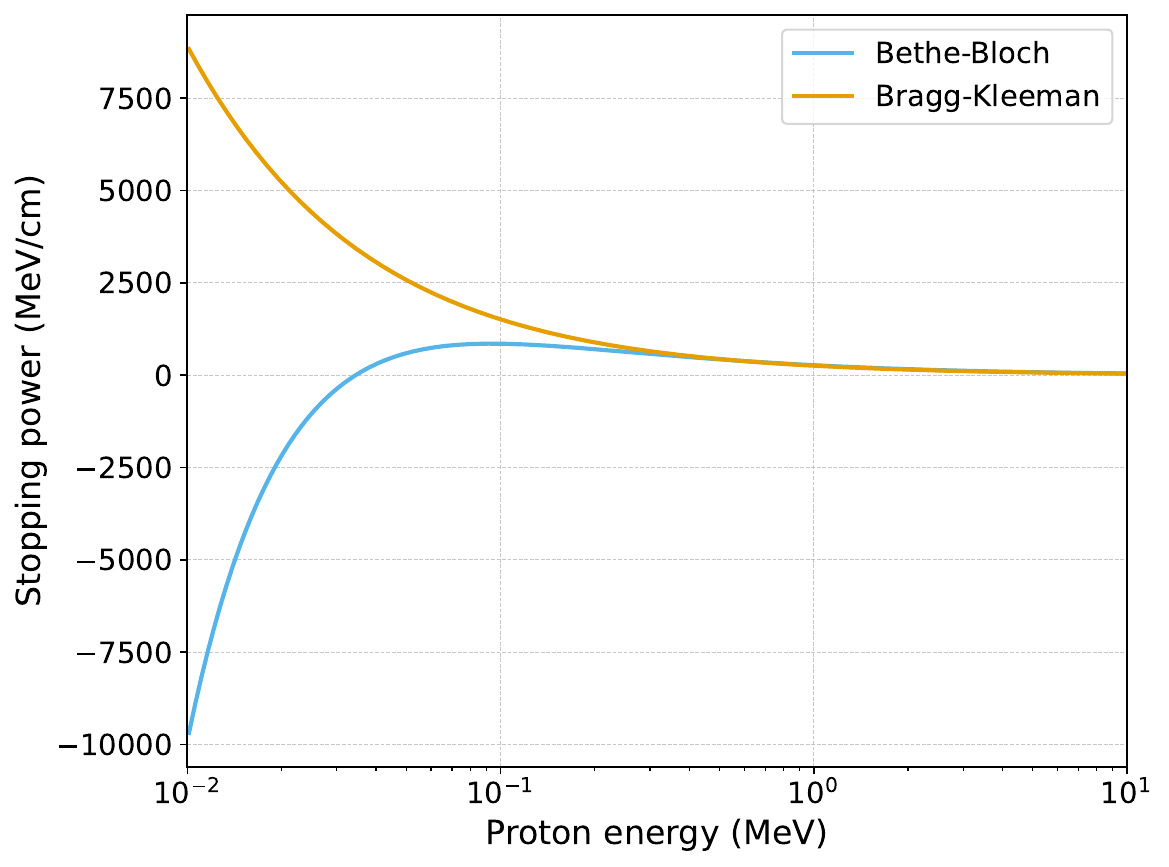}
    \caption{
      \label{fig:bethebloch}
      Stopping power as a function of energy for the Bragg-Kleeman and
      Bethe-Bloch models. The two models show good agreement in the
      intermediate energy range relevant to proton therapy however
      differ dramatically in the low energy range.}
  \end{figure}
\end{remark}

\subsection{Closed form solution.}

Due to the form of the stopping power, the equation is
hyperbolic in nature, we can therefore use the method of
characteristics to construct a closed form solution. To that end, the
characteristic curves satisfy
\begin{equation}
  E^p = E_{\max}^p - \frac{z}{\alpha}.
\end{equation}
This curve represents the energy trajectory of a monoenergetic proton
beam with initial energy $E_{\max}$ in the $(z, E)$-plane. Integrating
\eqref{eq:1dpde} along these characteristic curves, the solution for
fluence $u(z, E)$ is
\begin{equation}\label{eq:defn_fluence}
  \begin{split}
    u(z, E)
    &=
    \qp{E^p + \frac{z}{\alpha}}^{\frac{1-p}{p}}
    g\qp{\qp{E^p + \frac{z}{\alpha}}^{\frac{1}{p}}}
    E^{p-1}.
  \end{split}
\end{equation}

\subsection{Computation of Absorbed Dose}

The absorbed dose, $D(z)$, represents the energy deposited per unit
mass at a given depth and can be calculated by integrating the
stopping power weighted by the particle fluence over the energy
range 
\begin{equation}
  \label{eq:absorbed_dose}
  \begin{split}
    D(z)
    =&
    \int_{E_{\text{min}}}^{E_{\text{max}}}\dfrac{S(E)}{\rho(x)} u(z,E) \d E
    \\
    = &\frac{1}{\alpha p}\frac{1}{\rho(z)}
    \int_{E_{\text{min}}}^{E_{\text{max}}}
    \qp{E^p + \frac{z}{\alpha} }^{\frac{1-p}{p}}
    g\qp{\qp{E^p + \frac{z}{\alpha}}^{\frac{1}{p}}} \d E.
  \end{split}
\end{equation}
This formulation captures the full spectrum of proton energies within
$I$, accounting for their interactions with the medium and the
resulting energy deposition.

\begin{figure}[h!]
    \centering
    \includegraphics[width=\linewidth]{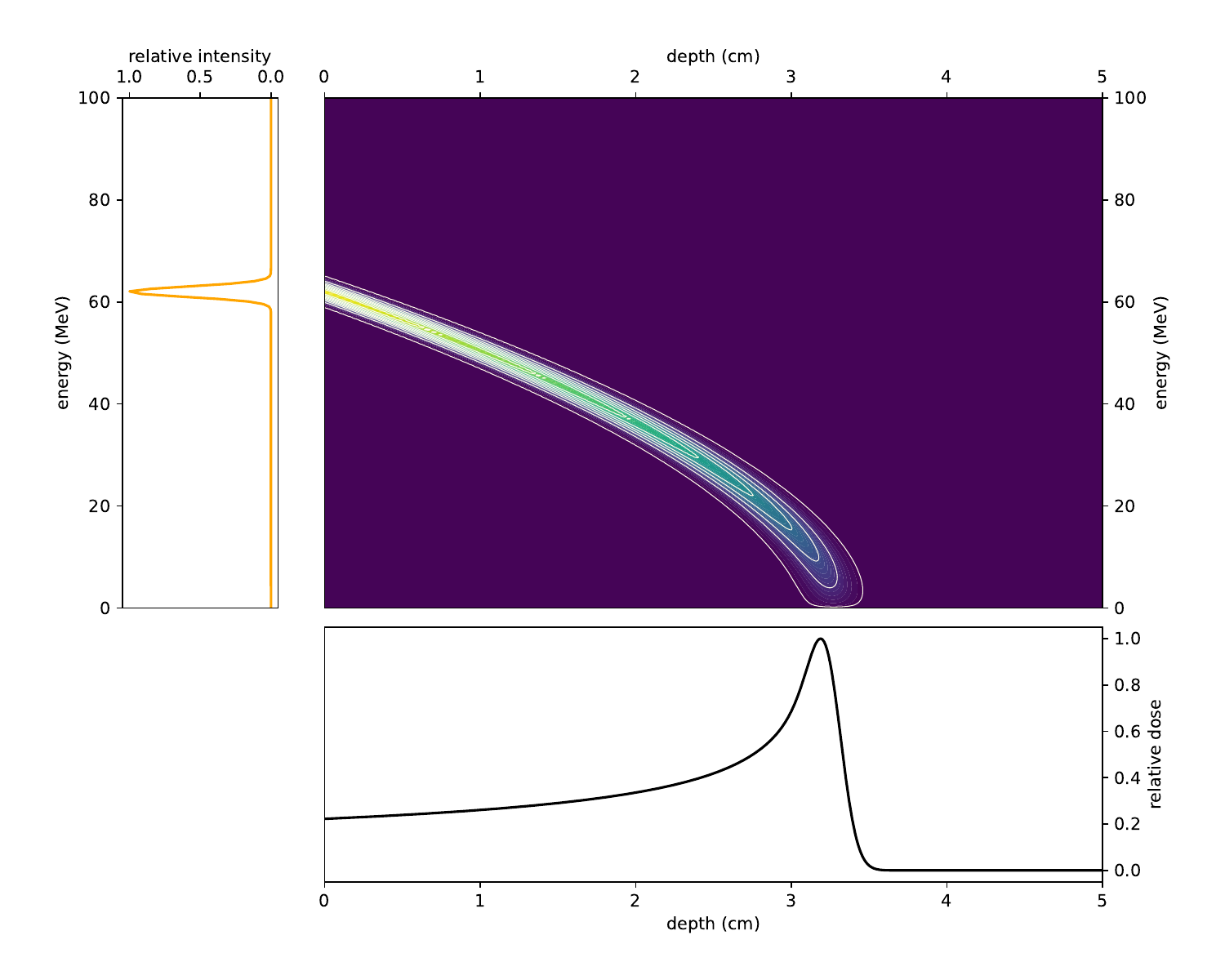}
    \caption{A visualisation of dose calculation. Left: the initial
      Gaussian energy profile of a 62 MeV proton beam with a 1\%
      energy spread. Middle: the fluence in depth-energy space,
      illustrating how the beam evolves as it travels through the
      medium. Bottom: the resultant dose as a function of depth.}
    \label{fig:prettyplot}
\end{figure}

To provide an intuitive understanding of this setup, Figure
\ref{fig:prettyplot} presents a visualisation of the fluence and
resulting dose for a $62$ MeV proton beam with a 1\% energy
spread. The figure consists of a grid of panels illustrating three
components:
\begin{itemize}
    \item The initial energy distribution $g(E)$ of the proton beam is
      shown on the left, highlighting the Gaussian profile centred at
      $62$ MeV.
    \item The middle panel depicts the fluence in depth-energy space,
      demonstrating how the protons’ energy evolves along their
      trajectories as they penetrate the medium.
    \item The bottom panel shows the resulting dose $D(z)$, plotted as
      a function of depth, capturing the energy deposited within the
      target region.
\end{itemize}
This visualisation effectively connects the initial beam properties to
the resulting dose distribution within the context of
\eqref{eq:absorbed_dose}.

\subsection{Linear Energy Transfer}

A related concept to stopping power is the Linear Energy Transfer
(LET), which describes the energy locally absorbed by the medium per
unit distance. The physical definition from the ICRU
\cite{thomas2012icru} defines LET, or the restricted linear electronic
stopping power as
\begin{equation}
    L_{\Delta}= \frac{\d E_{\Delta}}{\d z},
\end{equation}
which represents the energy lost $\d E_{\Delta}$ by the primary
charged particle in interactions with electrons, along a distance $\d
x$ minus energy carried away by energetic secondary electrons having
initial kinetic energies greater than $\Delta$. In the limit $ \Delta
\to \infty$, LET is equivalent to the stopping power, this is referred
to as the unrestricted LET. Essentially, stopping power accounts for
the total energy loss, while LET focuses on the energy absorbed
locally in the medium, which will be important in understanding the
biological impact of radiation later in this work.

If we assume that all of the energy lost is absorbed locally by the
material, then we have two notions of calculating the average LET from
the particle fluence: track-averaged and dose-averaged
\cite{kalholm2021systematic}.
\subsubsection{Track-averaged LET}
\begin{equation}\label{eq:Track_av_LET}
    L_T(z) 
    =
    \frac{\int_E u(z, E) S(E) \d E}{\int_E u(z, E) \d E}.
\end{equation}

\subsubsection{Dose-averaged LET}
\begin{equation}\label{eq:Dose_av_LET}
    L_D(z) 
    =
    \frac{\int_E u(z, E) S^2(E) \d E}{\int_E u(z, E)S(E) \d E}.
\end{equation}

\begin{remark}[Monoenergetic simplification]
  The track-averaged LET is sometimes referred to as fluence-averaged
  or particle-averaged LET. Track- and dose-averaged LET are defined
  for a polyenergetic beam, where $g(E)$ is an arbitrary distribution
  of energies. However, in the case of a truly monoenergetic beam,
  where
  \begin{equation}
    g(E) = c \delta(E - E_0)
  \end{equation}
  for some constant $c$ and initial energy $E_0 \in \mathbb{R}$, the
  expressions for track-averaged and dose-averaged LET simplify to
  \begin{equation}
    L_T(z)
    =
    L_D(z)
    =
    S\qp{
      \qp{E_0^p - \frac{z}{\alpha}}^{\frac{1}{p}}
    }.
  \end{equation}
\end{remark}

\subsection{Comparison with Monte Carlo Codes}

This section compares the analytical model described above with
established simulation tools to validate its performance against key
benchmarks. Specifically, we evaluate whether the simplified model,
given by \autoref{eq:1dpde}, can accurately reproduce both
qualitatively correct and quantitatively reasonable behaviour. The
comparison is conducted against the Monte Carlo simulation tool
MCsquare \cite{souris2016fast} and the Geant4-based TOPAS framework
\cite{faddegon2020topas}. 

To assess the accuracy of the one-dimensional analytical model, we
consider a pristine Bragg peak simulation as a computational
benchmark. A mono-energetic proton beam with an initial energy of $
E_0 = 62 \, \text{MeV}$ and a fluence of $1.21 \,
\text{gigaprotons/cm}^2$ is used as the input beam. The analytical
model employs standard Bragg-Kleeman parameter values for water ($
\alpha = 2.2 \times 10^{-3}$, $p = 1.77$) as reported in
\cite{bortfeld1997analytical}, while the Monte Carlo codes use their
respective default material parameters for water. To account for the
energy spread in the Monte Carlo simulations, the standard deviation
of the proton energy spectrum is set to $\epsilon E_0$, with $\epsilon
= 0.01$.

The boundary condition for the analytical model is defined as:
\begin{equation}\label{eq:gigaproton_spectrum}
  u(0, E) = 1.21 \times 10^9 \times C \exp\qp{-\frac{(E - E_0)^2}{2 \epsilon^2 E_0^2}},
\end{equation}
where $C$ is a normalisation constant ensuring that the integral of
the spectrum matches the total fluence of $1.21 \,
\text{gigaprotons/cm}^2$.

For the Monte Carlo codes, a three-dimensional water phantom is
simulated and the depth-dose curve is obtained by integrating the
dose over the plane perpendicular to the beam axis. Depth-dose curves
for the analytical model are computed using
\autoref{eq:absorbed_dose}.

\begin{figure}[h!]
    \centering
    \includegraphics[width=0.495\linewidth]{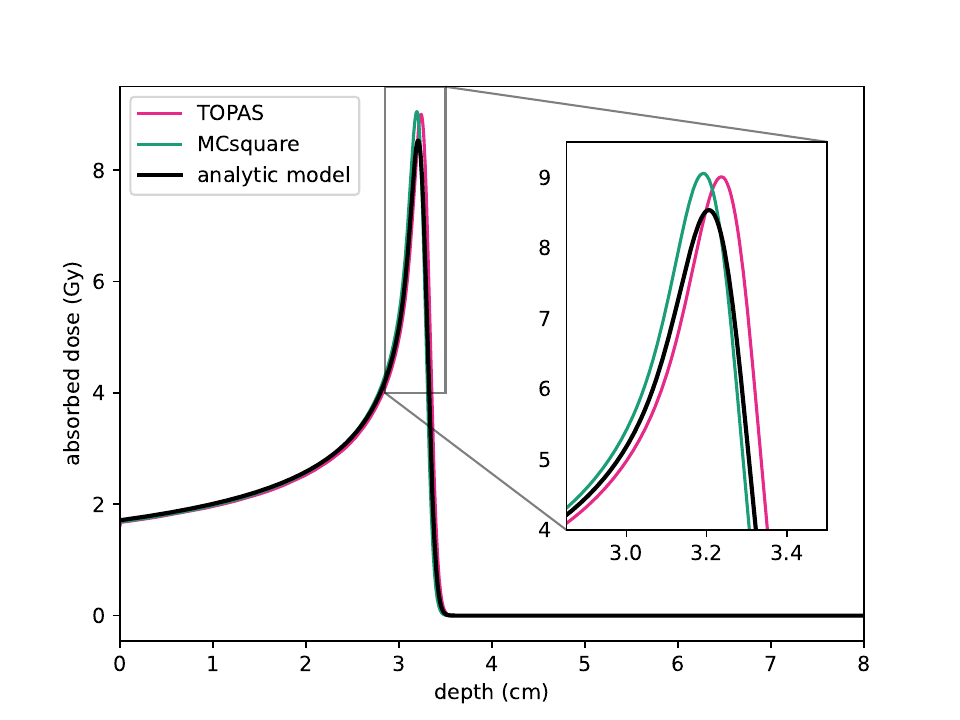}
    \includegraphics[width=0.495\linewidth]{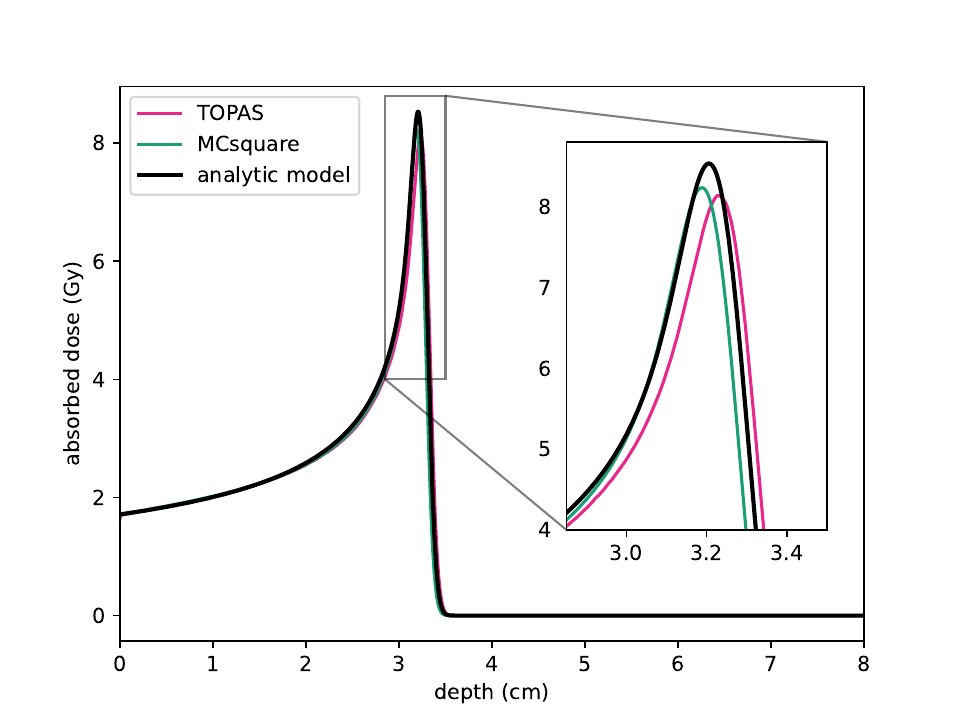}
    \caption{Comparison of depth-dose curves for a 62 MeV
      mono-energetic proton beam in water obtained from the
      one-dimensional analytical model (black), MCsquare (green) and
      TOPAS (pink). Left: nuclear interactions are excluded in the
      Monte Carlo simulations. Right: nuclear interactions are
      included. All simulations are scaled such that the total number
      of incoming protons is $1.21 \, \text{gigaprotons}$, ensuring a
      consistent comparison of absorbed dose.}
    \label{fig:compare_all_models}
\end{figure}

The comparison results are shown in \autoref{fig:compare_all_models},
where close agreement between the models can be observed. The left
panel shows results excluding nuclear interactions in the Monte Carlo
simulations, while the right panel includes these effects. Both cases
demonstrate that the analytical model captures the depth-dose
behaviour with high fidelity, providing a computationally efficient
alternative to Monte Carlo simulations.

\section{Biological Metrics}
\label{sec:biological}

Treatment planning can be thought of as translating a physician's
prescription into a set of parameters that define the radiotherapy
delivered to a patient \cite{kooy2015intensity}. The treatment plan
should be \lq optimal' in some sense, as defined in Section
\ref{sec:optimisation}. Typically, the goal is to deliver a dose that
closely matches a prescribed target dose profile using a suitable
metric. Absorbed dose is widely used for this purpose because it is
measurable and can be computed accurately in numerical simulations. In
addition, prescriptions from clinicians are given as a dose due to the
long history of use and gathered expertise in photon based
radiotherapy.  However, dose is only a proxy for biological effect and
there is no simple one-to-one relationship between absorbed dose and
biological outcomes \cite{nystrom2020treatment}. For example, LET and
cell type must also be considered, and for this reason, metrics beyond
dose have been incorporated into carbon ion beam treatment planning
(where LET plays a larger role than for protons) for over two decades
\cite{kanai1999biophysical, inaniwa2015reformulation,
  karger2017rbe}. In this section, we focus on cell survival rates as
a biological metric to assess treatment plan quality and introduce the
concept of the relative biological effectiveness (RBE) for proton
beams, as well as the biological dose (BD).

To formalise these ideas, specific examples of biological effect must
be considered. The choice of metric should align with the desired
clinical endpoints and be measurable or detectable
\cite{karger2017rbe}. Examples of such endpoints include tumour
control probability (TCP) and normal tissue complication probability
(NTCP), which are often estimated from in vivo experiments. For
simplicity, we focus on the cell survival fraction, a quantity
measurable in vitro via clonogenic assay \cite{serrano2023simple}. A
cell is considered \lq killed' or inactive if it is unable to
proliferate. This metric provides an objective measure of treatment
efficacy while avoiding the complexities of more comprehensive metrics
like TCP. However, while the survival fraction is relevant for tumour
control, it may not be an ideal measure for toxicity in healthy
tissues \cite{nystrom2020treatment}, which is an important
consideration in treatment planning.

For X-rays, the survival fraction $\mathcal{SF}_{\text{X-ray}}$ is
accurately modelled in vitro as a function of absorbed dose
$D_{\text{X-ray}}(x)$ using the \emph{linear-quadratic model}
\cite{kellerer1978generalized}:
\begin{equation}
  \label{eq:Survival_Fraction_photon}
  \mathcal{SF}_{\text{X-ray}}(z; D_{\text{X-ray}})
  =
  \exp\left(
    -c_{\text{X-ray}}(z) D_{\text{X-ray}}(z) - \beta_{\text{X-ray}}(z) D_{\text{X-ray}}(z)^2
  \right),
\end{equation}
where $c_{\text{X-ray}}(z)$\footnote{Often denoted as $\alpha$ in the
literature, but here denoted $c$ to avoid confusion with parameters in
the Bragg-Kleeman model.} and $\beta_{\text{X-ray}}(z)$ are model
parameters typically estimated through regression, with their
dependence on cell species captured via their spatial variation.  We
note that the linear quadratic model becomes less accurate in some
regimes \cite{hanin2010cell}, but is effective in many practical dose
ranges.

\begin{remark}[Interpretation of the parameters in the linear-quadratic model]\label{rem:double_strand_breaks}
As described in \cite{chadwick1973molecular}, the parameters in the
linear-quadratic model have physical interpretations. Cell death
following irradiation is primarily caused by DNA double-strand breaks,
either from a single particle interaction or from two single-strand
breaks created by separate interactions. The parameters
$c_{\text{X-ray}}(z)$ and $\beta_{\text{X-ray}}(z)$ correspond to the
expected number of single and double-strand breaks per unit dose,
respectively. While a general derivation of the model is given in
\cite{kellerer1978generalized}, the parameters are often determined
empirically by fitting the model to experimental data.
\end{remark}

\subsection{Linear-Quadratic Model for Protons}

Equation \eqref{eq:Survival_Fraction_photon} implies a one-to-one
relationship between absorbed dose and cell killing.
Predicting cell survival following irradiation with proton beams is
more complex than for X-rays due to the additional dependence on
LET. This dependence can be incorporated into the linear-quadratic
(LQ) framework and many studies have investigated how LET affects the
parameters of the LQ model \cite{hawkins1998microdosimetric,
  goodhead1992direct, belli1993inactivation}. LET-dependent models
remain widely used and well-studied \cite{wilkens2004phenomenological,
  chaudhary2014relative}.

Experimental evidence suggests that, within clinically relevant LET
ranges, the coefficient $c$ depends approximately linearly on LET:
\begin{equation}\label{eq:sf_with_let}
  c(z; L_D) = c_{0}(z) + \lambda(z) L_D(z),
\end{equation}
where $c_0(z)$ corresponds to the value of $c$ for X-rays and
$\lambda(z)$ represents the tissue-dependent LET sensitivity. A more
detailed model in \cite{tilly2002radiobiological} (see also
\cite{carabe2012range}) proposes modifying this dependence based on
the ratio of $c_{\text{X-ray}}$ to $\beta_{\text{X-ray}}$, introducing
exponential dependence when this ratio is large. However, for
simplicity, we assume $\lambda(z)$ is constant for a given tissue
type.

The relationship between $\beta$ and LET is less clear and constant
$\beta$ is commonly assumed \cite{chaudhary2014relative}. This
assumption aligns with the theory of dual radiation
action. Nevertheless, some studies, such as
\cite{carabe2007incorporation}, have examined potential LET-dependent
variations in $\beta$.

Following Chaudhary et al. \cite{chaudhary2014relative}, we set $c_0 =
c_{\text{X-ray}}$ in \autoref{eq:sf_with_let}. The spatial variation
of $c_{\text{X-ray}}$, $\beta$ and $\lambda$ reflects tissue-specific
responses to radiation. Table~\ref{tab:chaud} summarises these
parameters for two cell types: AG01522 (skin cells) and U87 (malignant
brain tumour cells) \cite{chaudhary2014relative}.
\begin{table}[h!]
\centering
\begin{tabular}{|l|c|c|c|}
\hline
Cell Type & $c_{\text{X-ray}}$ [$\text{Gy}^{-1}$] & $\lambda$ [$\mu \text{m}\, \text{keV}^{-1}\, \text{Gy}^{-1}$] & $\beta$ [$\text{Gy}^{-2}$] \\ 
\hline
AG01522 (skin) & $0.54 \pm 0.06$ & $0.0451$ & $0.051 \pm 0.038$ \\ 
\hline
U87 (brain tumour) & $0.11 \pm 0.028$ & $0.0127$ & $0.059 \pm 0.024$ \\ 
\hline
\end{tabular}
\caption{Parameters for the LET-dependent LQ model
  \eqref{eq:prot_survival}, measured for two cell types
  \cite{chaudhary2014relative}.}
\label{tab:chaud}
\end{table}

\begin{figure}
    \centering
    \includegraphics[width=0.45\linewidth]{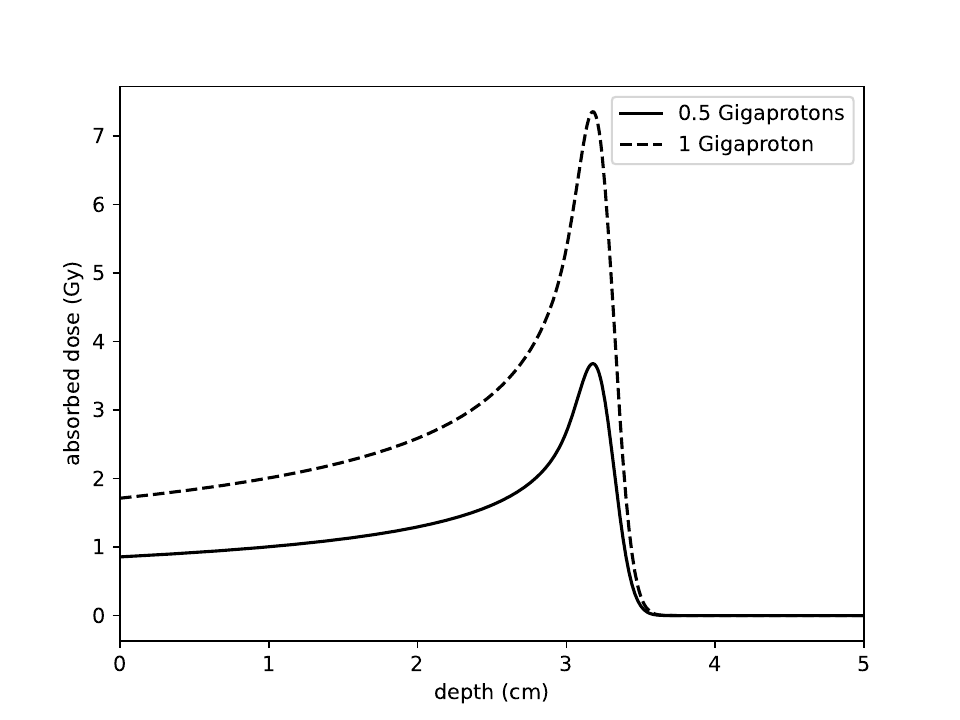}
    \includegraphics[width=0.45\linewidth]{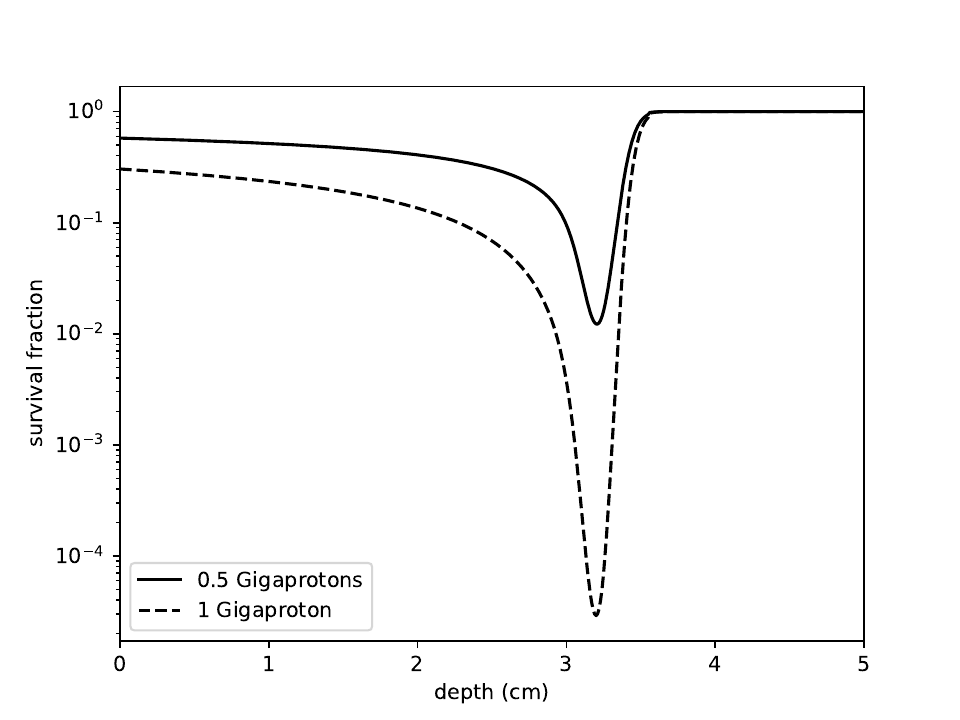}
    \caption{Left: depth-dose profiles for 62 MeV proton beams of two different intensities, Right: corresponding survival fractions of cells against depth assuming a homogeneous medium of cells, computed using the model \eqref{eq:prot_survival}.}
    \label{fig:bragg_peak_survival}
\end{figure}

The survival fraction of cells irradiated by a proton dose $D$ and LET
$L_D$ is then given by:
\begin{equation}\label{eq:prot_survival}
  \mathcal{SF}(z; D, L_D) = \exp\left(-c(z; L_D) \cdot D(z) - \beta(z) D(z)^2\right).
\end{equation}
For a given dose profile, $D(z)$, the surviving fraction of cells at
each depth $z$ may be predicted by the model
\eqref{eq:prot_survival}. An example for a Bragg peak is shown in
Figure \ref{fig:bragg_peak_survival}.  It is worth noting that this
model may require further modification for heavier charged particles
such as carbon ions \cite{wilkens2004phenomenological}.

\subsection{Relative Biological Effectiveness}
\label{sec:rbe}

Relative Biological Effectiveness (RBE) is an important metric in
radiobiology, defined as the ratio of doses required from two
radiation sources to achieve the same biological effect. It captures
differences in energy deposition and clinical outcomes between
radiation modalities \cite{paganetti2019report}. Here, we consider RBE
in terms of cell survival fraction.

Proton beams differ from X-rays in their energy deposition
characteristics. In particular, low-energy protons near the distal end
of their range inflict greater biological damage on tissue. In
\cite{hojo2017difference}, an increase in molecular markers of
double-strand DNA breaks ($\gamma$H2AX foci) is observed at the distal
end of the Bragg peak, consistent with
Remark~\ref{rem:double_strand_breaks}. It is well known that the RBE
of protons exceeds 1.0 and a constant value of 1.1 has been widely
adopted in clinical practice \cite{gerweck1999relative,
  underwood2016variable, giantsoudi2013linear,
  paganetti2019report}. While some studies support this approximation,
others suggest it is insufficient \cite{tilly2002radiobiological}.

Quantifying RBE is important to fully exploit the advantages of proton
therapy. However, this task is challenging due to complex dependencies
on biological factors, including cell type, cell cycle phase
\cite{underwood2016variable} and clinical endpoint
\cite{paganetti2019report}, as well as limited data for many tissues
\cite{grassberger2011variations}. Results from in vitro experiments on
specific cell lines often do not generalise to others due to
significant biological variation. Additionally, RBE depends on dose
and radiation quality, typically characterised by LET
\cite{wilkens2004phenomenological}.

\begin{figure}
    \centering
    \includegraphics[width=0.45\linewidth]{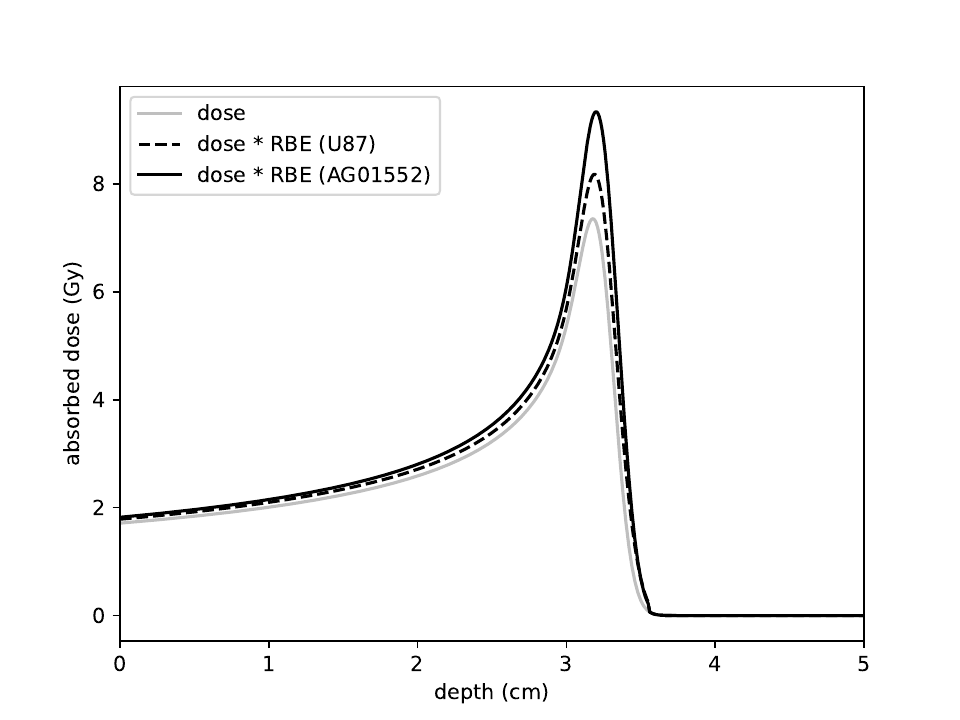}
    \includegraphics[width=0.45\linewidth]{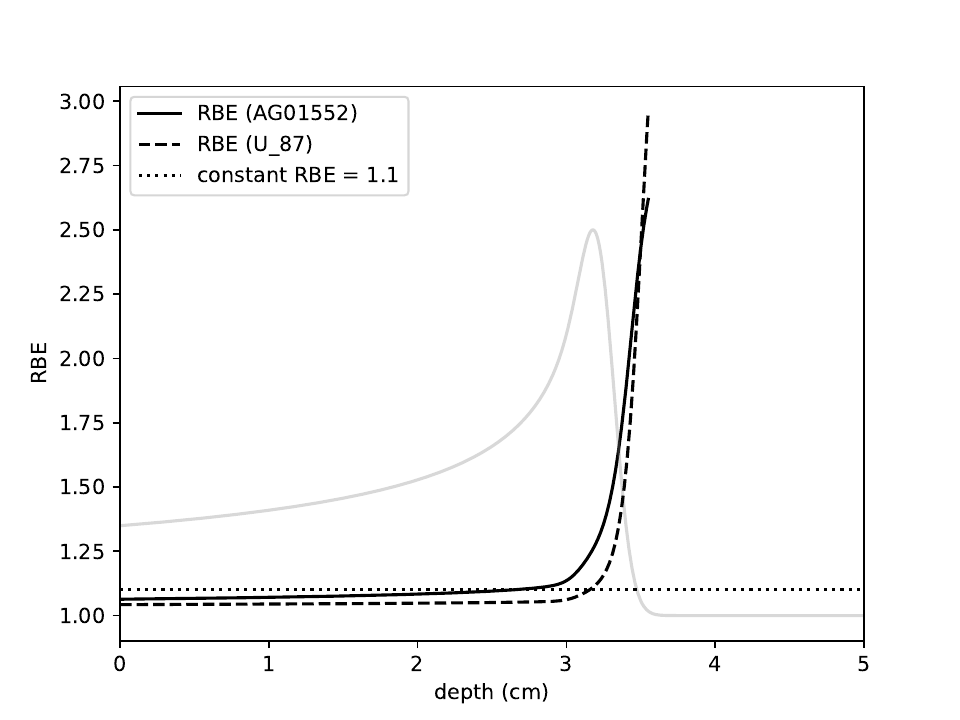}
    \caption{Left: RBE-weighted dose curves for a 62 MeV
      mono-energetic proton beam in water phantom, calculated using
      the TDRA model for cell survival and parameters from
      \cite{chaudhary2014relative} for AG01522 and U87 cell
      lines. Right: Corresponding RBE. Dose curve illustrates RBE
      behaviour along the Bragg peak and is not to scale.}
    \label{fig:rbes_for_two_cell_lines}
\end{figure}

In this work, we consider a spatially variable notion of relative
biological effectiveness as follows. Given a depth dose curve $D(z)$
and LET profile $L_D(z)$, at each point $z$ a surviving fraction of
cells may be computed as described above.  The equivalent photon dose,
i.e. the dose that yields the same cell survival fraction, is found by
inverting the functional relationship given in Equation
\eqref{eq:Survival_Fraction_photon}. Specifically, given $D$, we seek
$D_{\text{X-ray}}$ such that:
\begin{equation}
  \mathcal{SF}(z; D(z), L_D(z)) = \mathcal{SF}_{\text{X-ray}}(z; D_{\text{X-ray}}(z))
\end{equation}
for all $z$. We may then define a spatially variable relative
biological effectiveness as
\begin{equation}
  RBE(z, D, L_D)
  :=
  \frac{D_{\text{X-ray}}(z)}{D(z)}.
\end{equation}
The RBE-weighted dose is then defined to be the product of dose and
RBE, i.e. precisely the equivalent X-ray dose $D_{\text{X-ray}}$. This
quantity can be used to compare the biological effect with photon dose
curves. RBE and RBE-weighted dose curves are shown in Figure
\ref{fig:rbes_for_two_cell_lines} for two different cell types studied
in \cite{chaudhary2014relative}. We observe that the RBE is in line
with the clinical value of 1.1 up until the Bragg peak, but becomes
significantly larger in the distal fallof region.

\subsection{Biological Dose (BD)}

Unfortunately, optimising for dose delivery that results in a given
survival fraction is problematic for a number of reasons. Discussion
of these issues is postponed until \S\ref{sec:optimisation}. In this
section an alternative is presented which is more amenable to
optimisation.

A natural alternative is to consider the logarithm of the survival
fraction, as is common in log-likelihood maximisation. This idea
appears in \cite{unkelbach2016reoptimization, mcintyre2023systematic},
where the LET-weighted dose, often referred to as the biological dose
(BD), is used as an optimisation metric. In \cite[Appendix
  A]{unkelbach2016reoptimization}, a discussion of a simpler linear
exponential model for the survival fraction is used is given. It can
be viewed as a simplification of the linear quadratic model for which
fractionation effects are neglected. The model is given by:
\begin{equation}
    \mathcal{SF}_{\text{lin}}(z; D(z), L_D(z))
    :=
    \exp\left(- c(z; L_D) D(z)\right).
\end{equation}
The biological dose is then defined as 
\begin{equation}
    BD(z)
    :=
    - \log(\mathcal{SF}_{\text{lin}}(z; D(z), L_D(z)))/ c_{\text{X-ray}}
    =
    D(z) \left(1 + \frac{\lambda}{c_{\text{X-ray}}} L_D(z)\right),
\end{equation}
where $D(z)$ is the absorbed dose, $L_D(z)$ is the dose-averaged LET,
and $\lambda / c_{\text{X-ray}}$ quantifies the contribution of LET to
the biological effect (cf. Equations \eqref{eq:sf_with_let} \&
\eqref{eq:prot_survival}). This formulation balances physical dose
delivery with biological considerations, making it a promising metric
for treatment planning.

\section{Model Uncertainties and Sensitivity}
\label{sec:uncertainties} 

This section aims to quantify the uncertainty in the magnitude and
position of $D(z)$ \eqref{eq:absorbed_dose} when the stopping power
parameters $\alpha$ and $p$ from the Bragg-Kleeman rule in
\autoref{eq:BraggKleeman} are uncertain. Specifically, we consider the
dose as parameterised by $\alpha$ and $p$, such that $D(z) =
D(z;\alpha,p)$. Two methods, an active subspace approach and a Monte
Carlo simulation, are implemented to analyse the impact of uncertainty
in the model.

\subsection{Active Subspace Method} \label{sec:active_subspace_method}

We apply an active subspace method to evaluate the relative importance
of the parameters $\alpha$ and $p$ over specified ranges. This
approach involves examining the dose across the phase space of
$\alpha$ and $p$ at a fixed point in the domain. The direction
perpendicular to the contour lines in this space indicates the path
along which the greatest change in dose occurs, providing insight into
the relative sensitivities of $D(z;\alpha,p)$ to $\alpha$ and
$p$. Furthermore, the orientation of these contour lines is orthogonal
to the eigenvectors of the covariance matrix at that point (see
\cite[\S 10.5]{Sullivan-2015} for further details).

The parameters $\alpha$ and $p$ are modelled as independent normally
distributed random variables, with means $(\mu_\alpha, \mu_p)$ and
variances $(\sigma_\alpha, \sigma_p)$:
\begin{equation}
  \label{eq:alpha_p_normal}
  \begin{split}
    \alpha &\sim \mathcal{N}(\mu_\alpha, \sigma_\alpha)
    \\
    p &\sim \mathcal{N}(\mu_p, \sigma_p).
  \end{split}
\end{equation}
    For all simulations, we take $\mu_\alpha = 0.00246$ and $\mu_p =
1.75$, as shown in Table \ref{tab:range_energy}.

We consider three cases for the standard deviations $\sigma_\alpha$
and $\sigma_p$: absolute values, relative values, and empirical
estimates based on data. For the absolute case, we set $\sigma_\alpha
= \sigma_p = 0.0001$. For the relative case, $\sigma_\alpha$ and
$\sigma_p$ are taken to be 1\% of their respective mean values. For
the empirical case, we use the data from Table \ref{tab:range_energy},
assuming the 95\% confidence intervals represent twice the standard
deviation. Thus, $\sigma_\alpha$ and $\sigma_p$ are scaled by $1/1.96$
to align the normal distribution with these intervals. Table
\ref{tab:active_subspace_choices} summarises these choices.

\begin{table}[h!]
    \centering
    \begin{tabular}{|c|c|c|}
    \hline
    Case & $\sigma_\alpha$ & $\sigma_p$ \\
    \hline
    Absolute & 0.0001 & 0.0001 \\
    Relative & 0.0000246 & 0.0175 \\
    Empirical & 0.000128 & 0.0102 \\
    \hline
    \end{tabular}
    \caption{Standard deviations $\sigma_\alpha$ and $\sigma_p$ for the three case studies.}
    \label{tab:active_subspace_choices}
\end{table}

\begin{figure}[h!]
    \centering
    \includegraphics[width=0.5\linewidth]{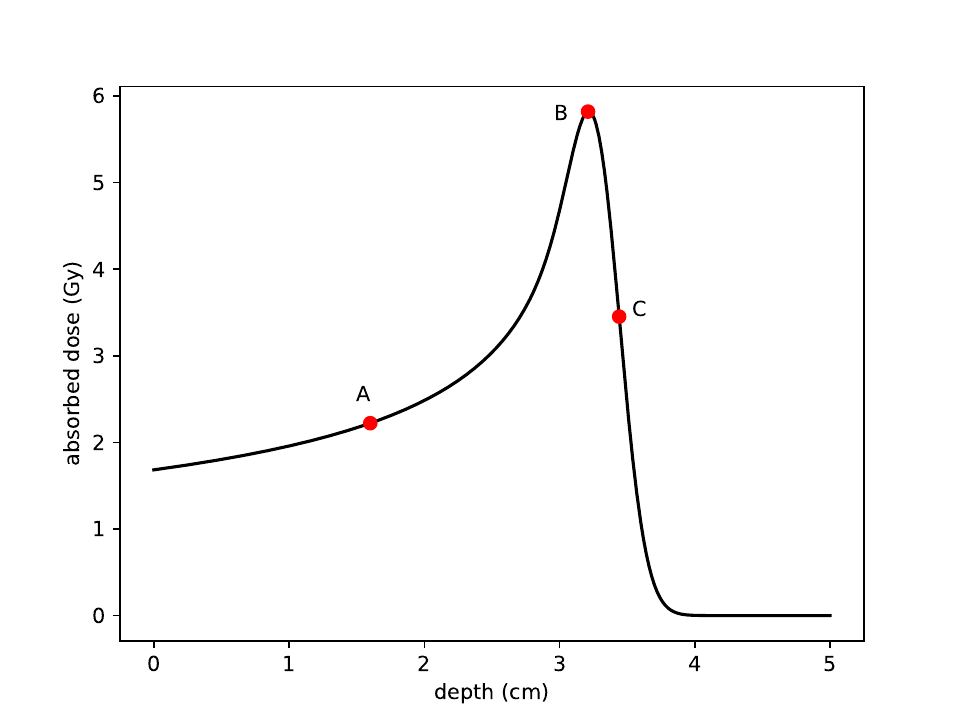}
    \caption{Positions for the active subspace analysis. Point $A$ is
      halfway between the start and the Bragg peak, point $B$ is at
      the peak, and point $C$ is at the position with the steepest
      gradient. The initial beam has an energy of 62MeV, with a spread of $5\%$.}
    \label{fig:positions}
\end{figure}

A sensitivity analysis is conducted at three points in the domain,
labelled $A$, $B$, and $C$, as shown in Figure
\ref{fig:positions}. Point $A$ lies midway between the start of the
beam and the Bragg peak, point $B$ is at the Bragg peak, and point $C$
is at the location of the steepest gradient.

\begin{figure}[h!]
    \centering
\includegraphics[width=0.3\linewidth]{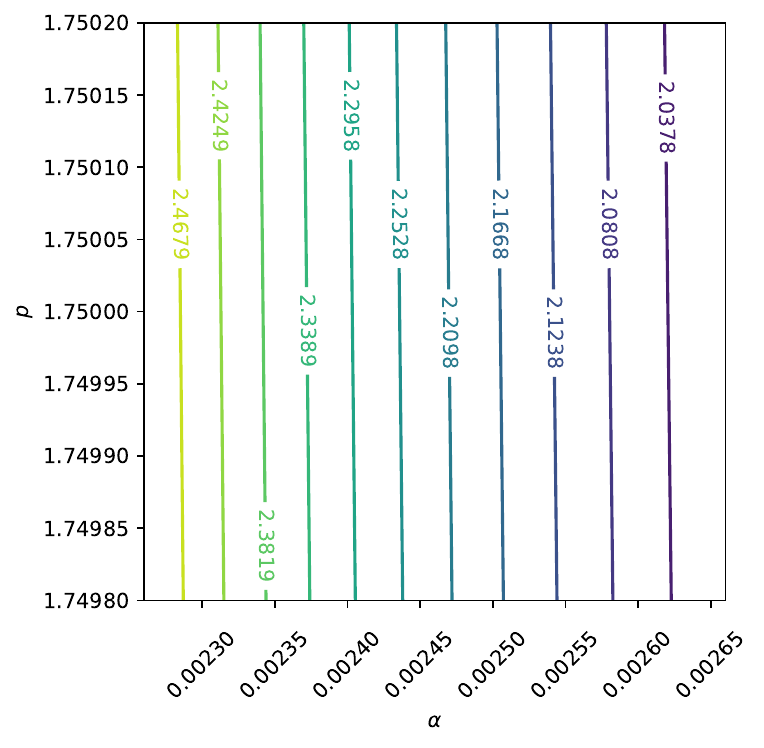}
\includegraphics[width=0.3\linewidth]{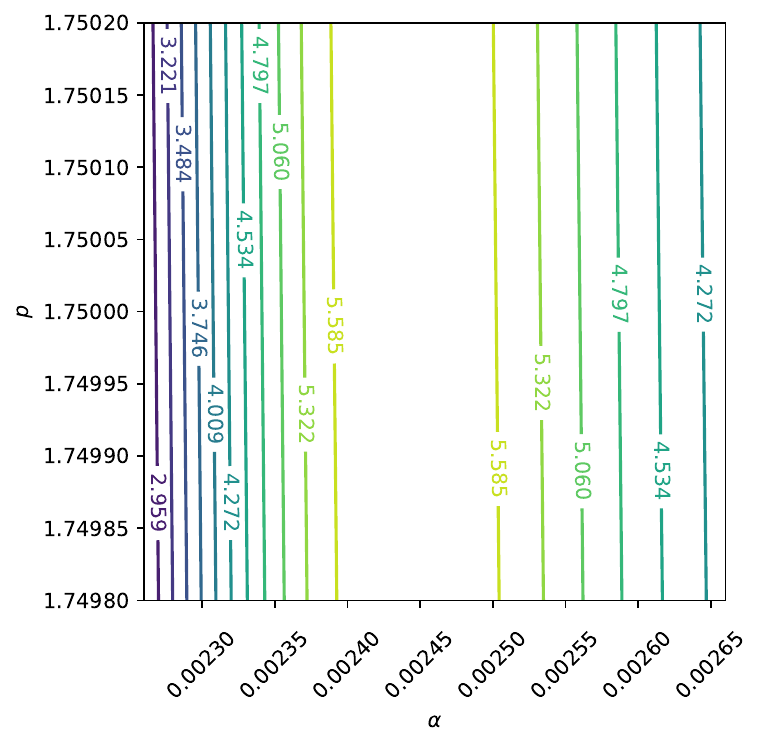}
\includegraphics[width=0.3\linewidth]{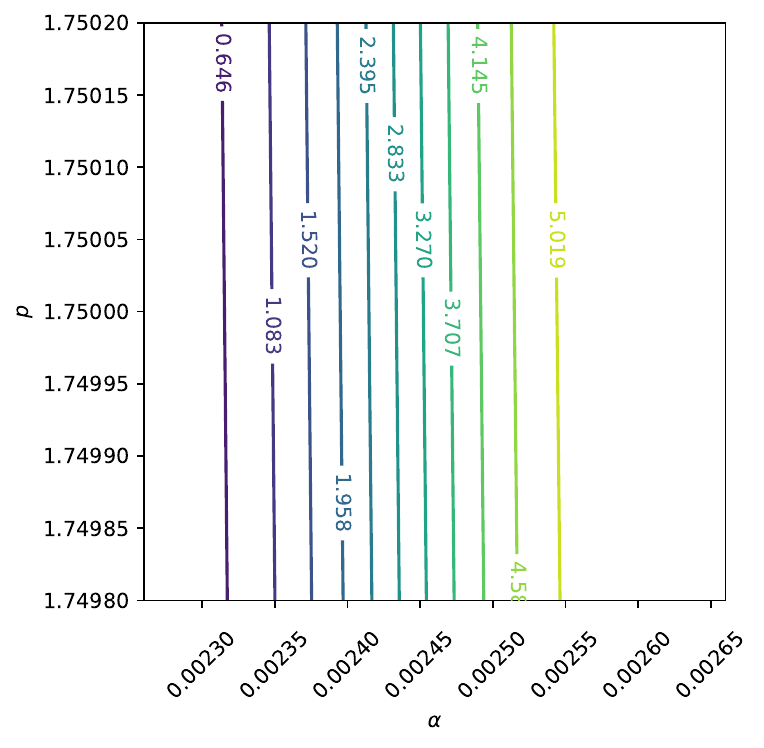}
\caption{Contour plot of dose at points A, B, and C when $\alpha$
  and $p$ are assumed to follow normal distributions with absolute
  standard deviations $\sigma_\alpha = \sigma_p = 0.0001$.}
    \label{fig:active_subspace2}
\end{figure}

\begin{figure}[h!]
    \centering
\includegraphics[width=0.3\linewidth]{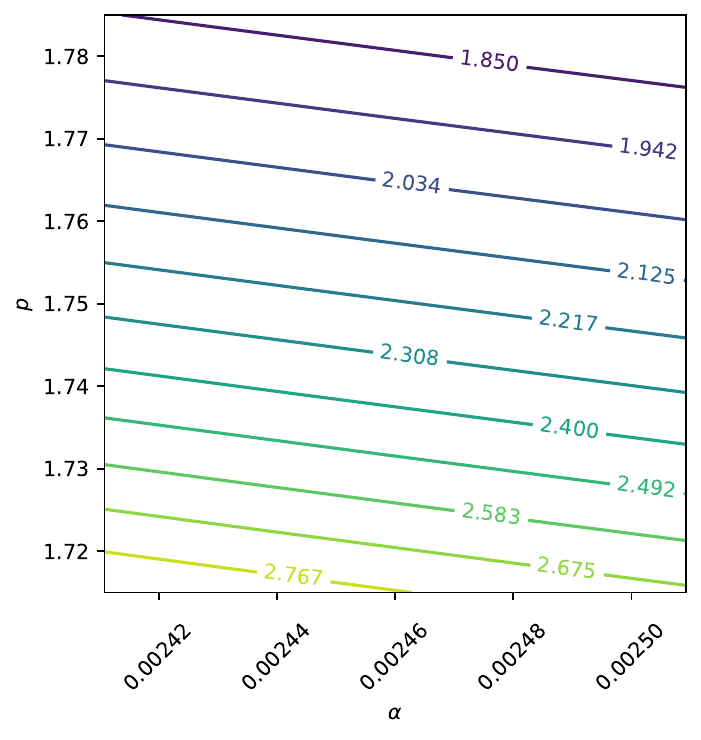}
\includegraphics[width=0.3\linewidth]{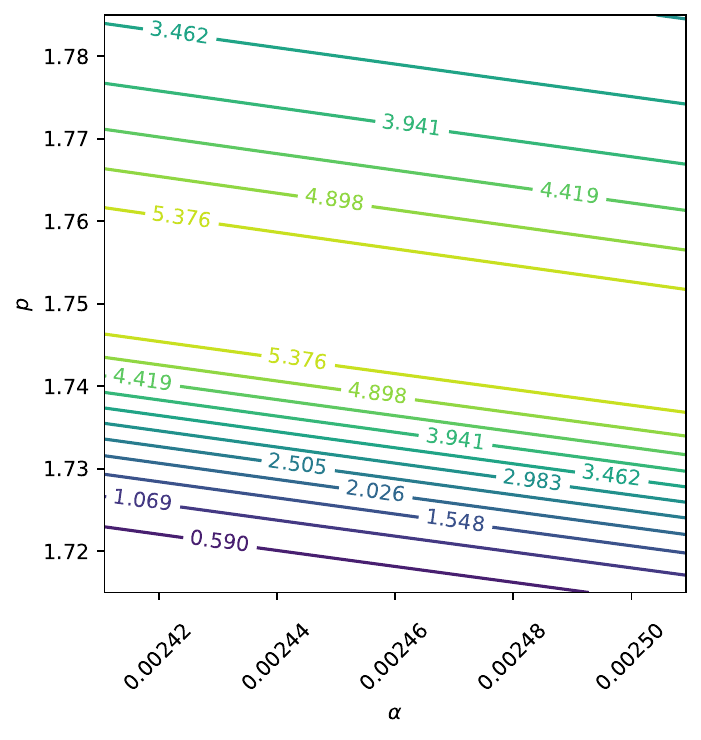}
\includegraphics[width=0.3\linewidth]{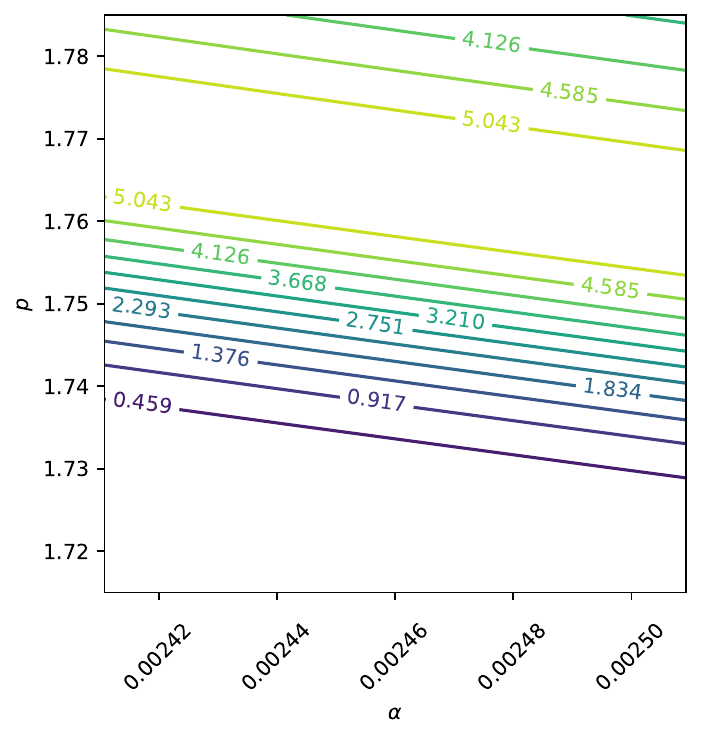}
\caption{Contour plot of dose at points A, B, and C when $\alpha$
  and $p$ are assumed to follow normal distributions with relative
  standard deviations $\sigma_\alpha = 0.01\mu_\alpha =
  0.0175$ and $\sigma_p = 0.01\mu_p = 0.0000246$.}
    \label{fig:active_subspace3}
\end{figure}

\begin{figure}[h!]
    \centering
\includegraphics[width=0.3\linewidth]{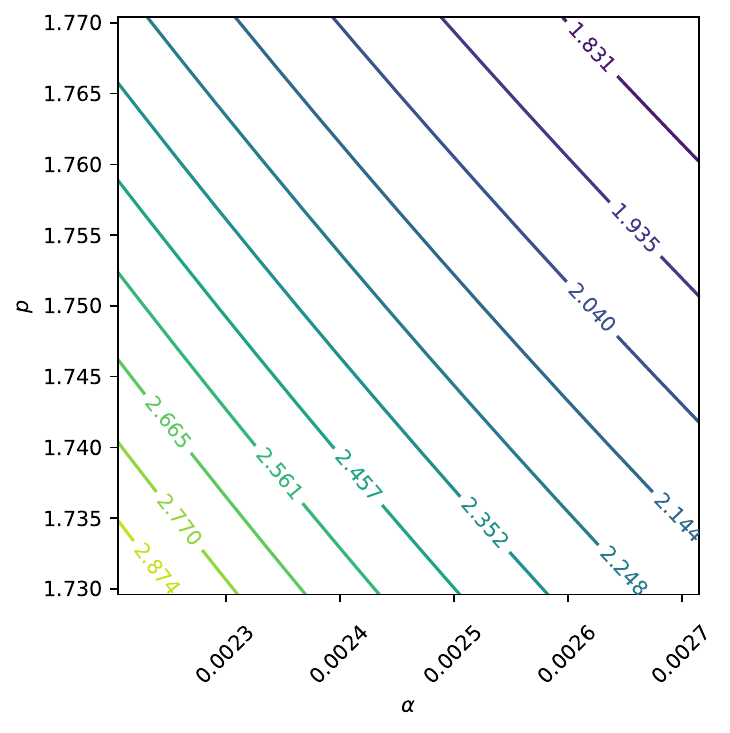}
\includegraphics[width=0.3\linewidth]{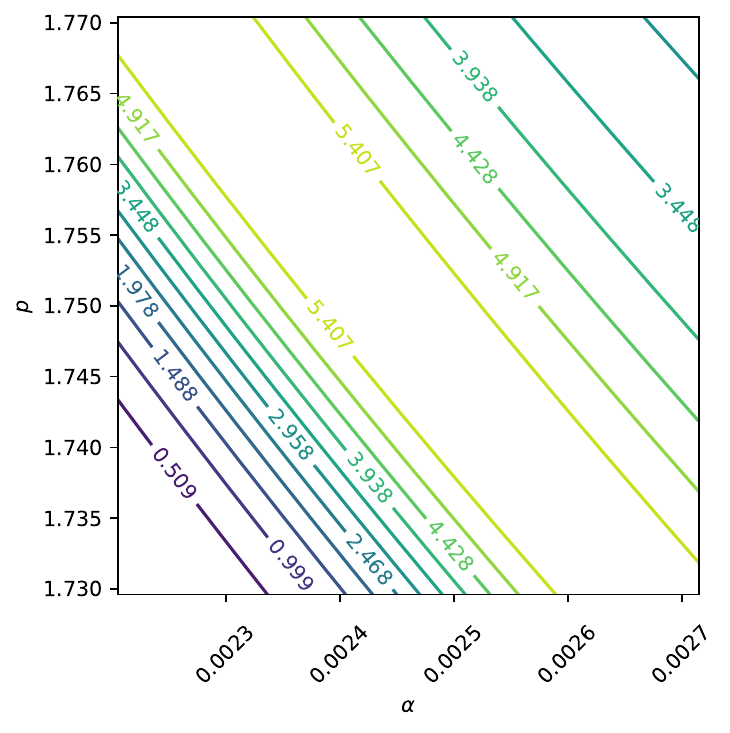}
\includegraphics[width=0.3\linewidth]{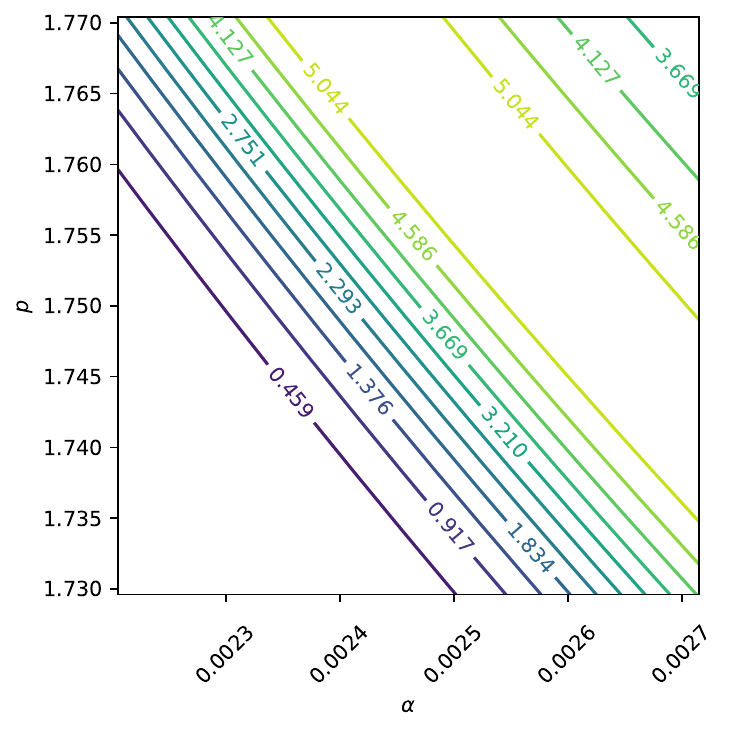}
\caption{Contour plot of dose at points A, B, and C when $\alpha$
  and $p$ are assumed to follow normal distributions with empirical
  standard deviations $\sigma_\alpha = 0.000128$ and $\sigma_p =
  0.0102$.}
    \label{fig:active_subspace1}
\end{figure}

Contour plots illustrating the sensitivities at points $A$, $B$, and
$C$ are shown in Figures \ref{fig:active_subspace2},
\ref{fig:active_subspace3}, and \ref{fig:active_subspace1},
respectively. The $x$- and $y$-axes of these figures represent the
ranges $(\mu_\alpha - 2\sigma_\alpha, \mu_\alpha + 2\sigma_\alpha)$
and $(\mu_p - 2\sigma_p, \mu_p + 2\sigma_p)$, ensuring that $\alpha$
and $p$ span their 95\% confidence intervals.

Figure \ref{fig:active_subspace2} illustrates that, when absolute
uncertainties are considered, the contour lines are nearly vertical,
indicating that $D(x;\alpha, p)$ is significantly more sensitive to
changes in $\alpha$ than to equivalent changes in $p$. In contrast,
Figure \ref{fig:active_subspace3} shows that when relative
uncertainties are used, a 1\% change in $p$ has a greater impact on
$D(x;\alpha, p)$ than a 1\% change in $\alpha$. Finally, Figure
\ref{fig:active_subspace1}, which uses empirical standard deviations
based on \autoref{tab:range_energy}, demonstrates that $D(x;\alpha,
p)$ is equally sensitive to both parameters over the examined phase
plane.

It is important to note that while \autoref{fig:active_subspace2}
highlights $\alpha$'s greater influence under absolute uncertainties
and \autoref{fig:active_subspace3} shows $p$'s dominance under
relative uncertainties, \autoref{fig:active_subspace1} reflects a
balance in sensitivity when using empirically motivated standard
deviations. This motivates the inclusion of uncertainty in both
$\alpha$ and $p$ in the subsequent Monte Carlo analysis.

\subsection{Monte Carlo Simulation}

In this section, we investigate how uncertainty in the stopping power
parameters $\alpha$ and $p$ affects the shape and position of the dose
curve $D(z;\alpha,p)$. Using Monte Carlo simulations, we quantify the
overall uncertainty in $D(z;\alpha,p)$ as well as the uncertainty in
the depth of the Bragg peak. As in
\autoref{sec:active_subspace_method}, we model $\alpha$ and $p$ as
independent, normally distributed random variables with means
$(\mu_{\alpha},\mu_p)$ and standard deviations
$(\sigma_{\alpha},\sigma_p)$. Specifically, we take $\mu_{\alpha} =
0.00246$ and $\mu_p = 1.75$, and use empirical standard deviations
$\sigma_{\alpha} = 0.000128$ and $\sigma_p = 0.0102$, corresponding to
$95\%$ confidence intervals.

The nominal dose curve, computed with $\alpha = \mu_{\alpha}$ and $p =
\mu_p$, is denoted as $D^*(z;\mu_{\alpha},\mu_p)$ and serves as the
``true'' reference dose curve. This curve corresponds to the assumed
parameter values $\mu_{\alpha}=0.00246$ and $\mu_{p}=1.75$, as
presented in \autoref{tab:range_energy}. In the figures that follow,
$D^*(z;\mu_{\alpha},\mu_p)$ is included to illustrate how
uncertainties in $\alpha$ and $p$ influence the dose curve.

It is important to note that the nominal dose curve
$D^*(z;\mu_{\alpha},\mu_p)$ is not equivalent to the mean dose curve,
as $D^*(z;\mathbb{E}[\alpha],\mathbb{E}[p]) \neq
\mathbb{E}[D(z,\alpha,p)]$. As a result, the nominal dose curve does
not necessarily lie within the calculated confidence intervals, which
reflect the distribution of $D(z;\alpha,p)$. This underscores the
impact of parameter uncertainty on the dose curve and highlights the
importance of considering the full range of variability in $\alpha$
and $p$.

\begin{figure}[h!]
    \centering
\includegraphics[width=0.3\textwidth]{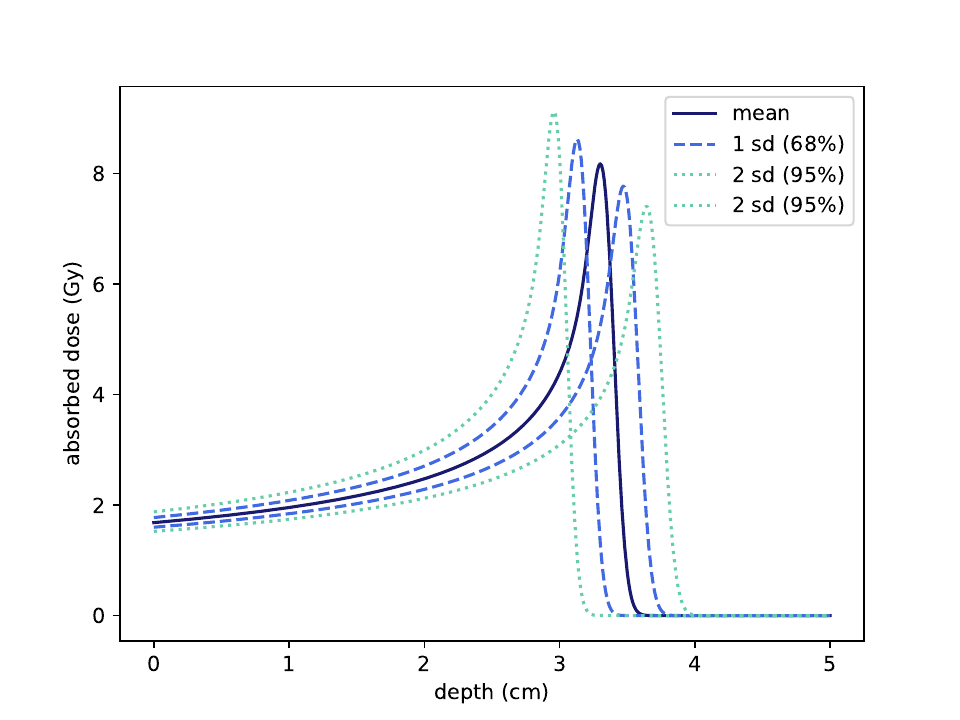}
\includegraphics[width=0.3\textwidth]{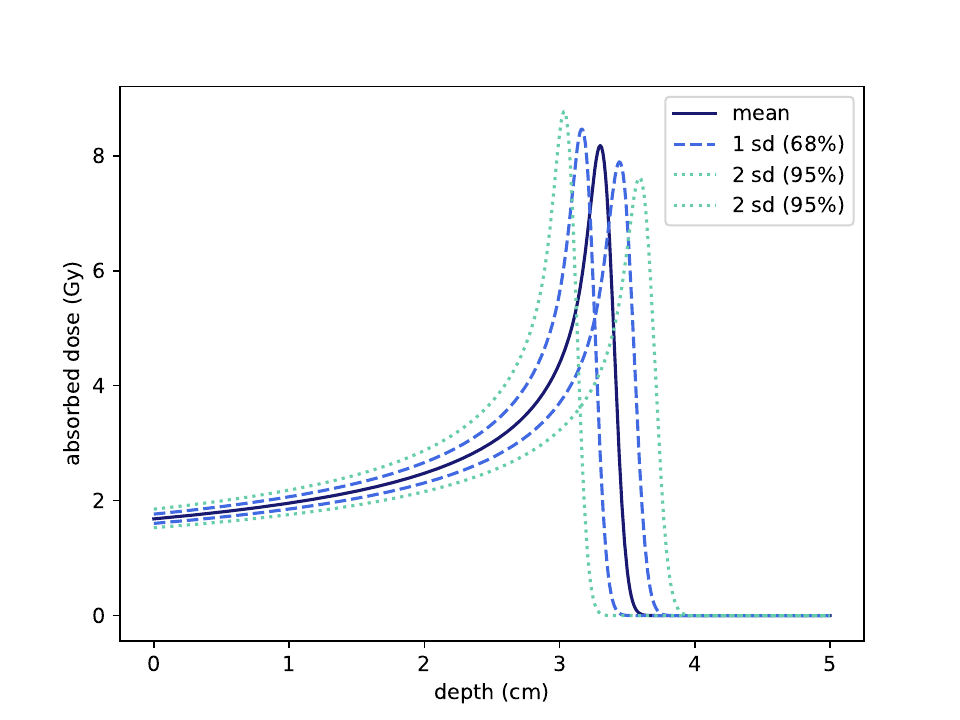}
\includegraphics[width=0.3\textwidth]{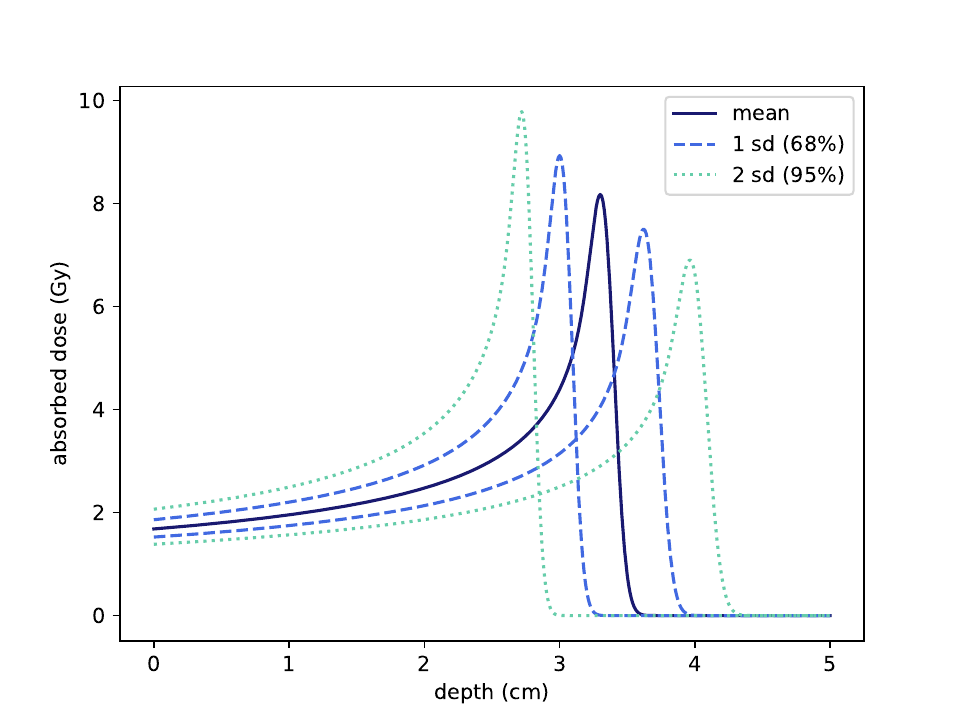}
\caption{Sensitivity of $D(z;\alpha,p)$ to under- and overestimation
  of $\alpha$ and $p$ by one or two standard deviations. The
  nominal dose curve $D^*(z;\mu_{\alpha},\mu_p)$ is shown for
  reference. On the left, only $\alpha$ has been under- or overestimated; in the centre only $p$; and on the right both $\alpha$ and $p$.
  }
    \label{fig:shifted_coefs}
\end{figure}

In \autoref{fig:shifted_coefs}, we show how the dose curve
changes when $\alpha$ and $p$ deviate by one or two standard
deviations in either the positive or negative direction. The results
indicate that deviations in either $\alpha$ or $p$ alone induce
moderate changes in $D(z;\alpha,p)$. However, simultaneous deviations
in both parameters cause larger changes in both the magnitude and the
location of the Bragg peak.

Notably, the probability of both $\alpha$ and $p$ being off by
$2\sigma$ simultaneously is significantly lower than the probability
of a single parameter deviating by $2\sigma$, assuming
independence. For instance, under the normal distribution assumption,
the probability of $\alpha$ being off by $2\sigma_{\alpha}$ is $5\%$,
while the probability of simultaneous $2\sigma$ deviations for both
parameters is $0.25\%$.

\begin{figure}[h!]
    \centering
    \includegraphics[width=0.3\linewidth]{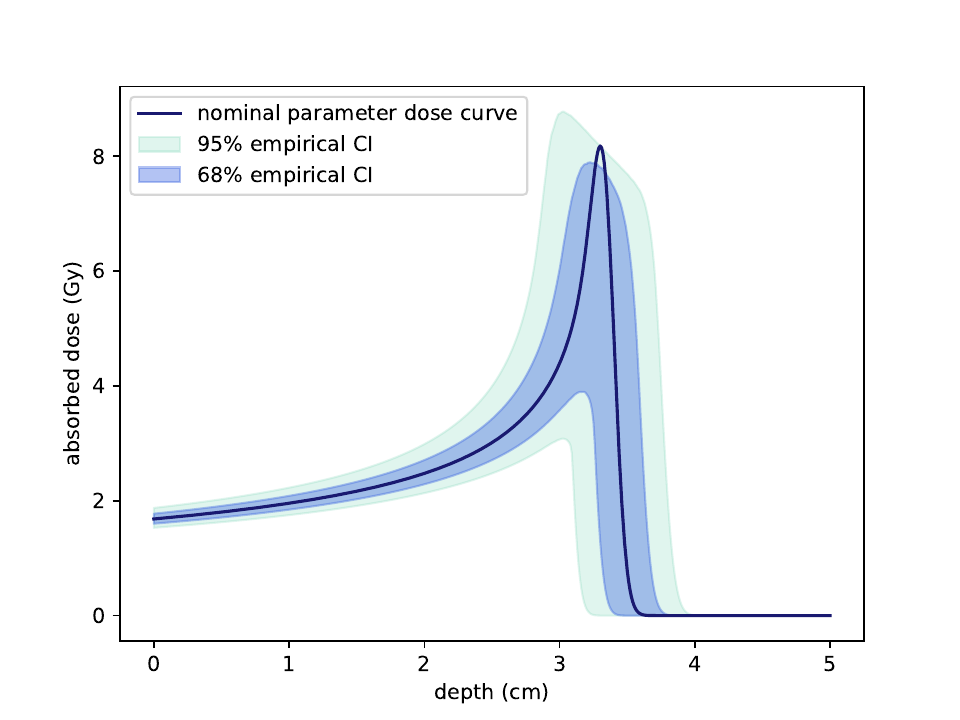}
    \includegraphics[width=0.3\linewidth]{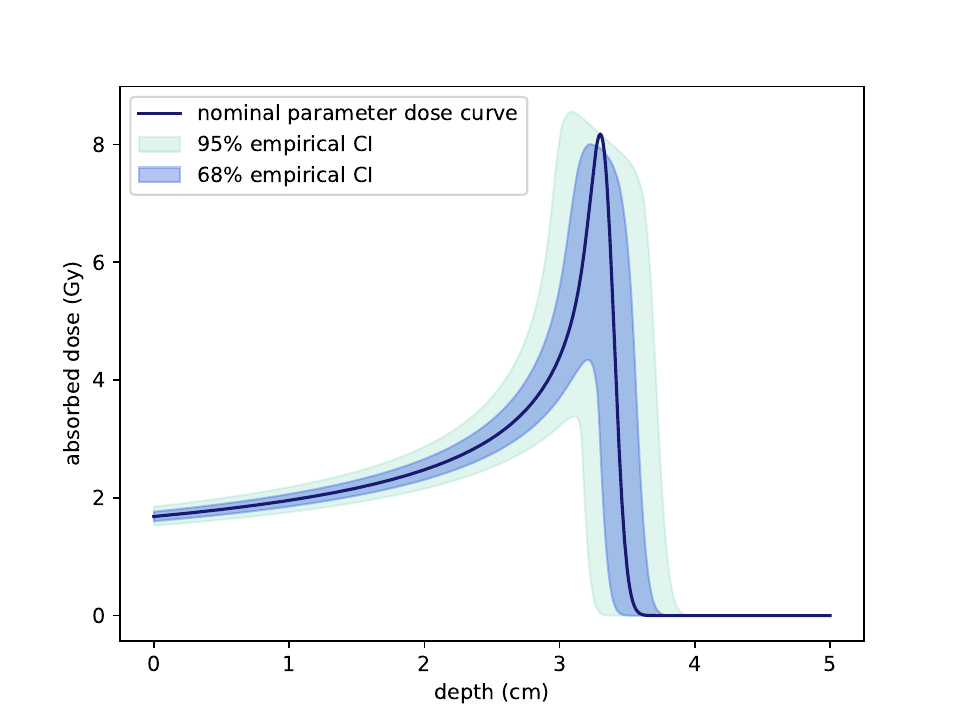}
    \includegraphics[width=0.3\linewidth]{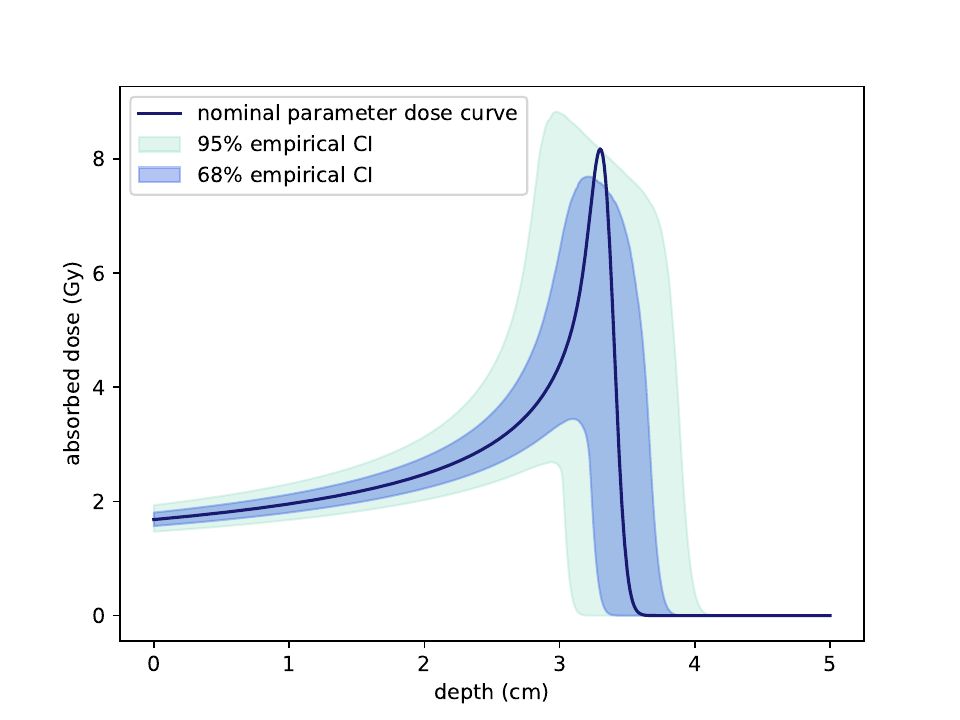}
    \caption{Confidence intervals for the variation in dose
      $D(z;\alpha,p)$ when $\alpha$ and $p$ are normally distributed
      with means $\mu_{\alpha} = 0.00246, \mu_p = 1.75$ and standard
      deviations $\sigma_{\alpha} = 0.000128, \sigma_p = 0.0102$. The
      nominal dose curve $D(z;\mu_{\alpha},\mu_p)$, resulting from the assumed parameter values $\mu_{\alpha}$ and $\mu_p$, is plotted in
      dark blue. The $68\%$ and $95\%$ confidence intervals are shown
      as shaded regions, estimated using 25,000 independent random
      samples of $\alpha$ and $p$. On the left, only uncertainty in $\alpha$ has been included; in the centre only uncertainty in $p$, and on the right uncertainty in both $\alpha$ and $p$.}
    \label{fig:dose_curve_ci}
\end{figure}

\autoref{fig:dose_curve_ci} presents the estimated $68\%$ and $95\%$
confidence intervals for the dose curve $D(z;\alpha,p)$. These
intervals are derived using a Monte Carlo approach, with 25,000
independent random samples of $\alpha$ and $p$. Empirical quantiles
are computed at each depth $z$ to generate the ensemble of dose
curves. 

The results show that introducing uncertainty in either $\alpha$ or
$p$ individually leads to similar confidence intervals. However, when
uncertainty is included for both parameters simultaneously, the
confidence intervals for the dose curve become significantly
larger. This demonstrates that the combined uncertainties in $\alpha$
and $p$ amplify the overall uncertainty in dose deposition,
emphasising the importance of accurately characterising both
parameters.

\begin{figure}[h!]
    \centering
    \includegraphics[width=0.3\linewidth]{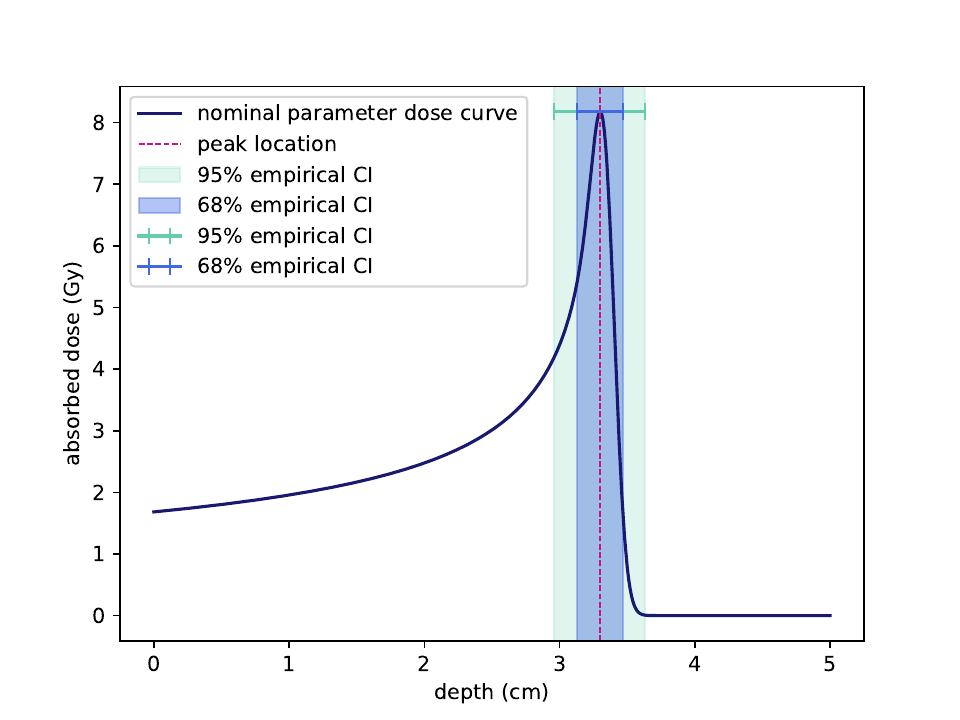}
    \includegraphics[width=0.3\linewidth]{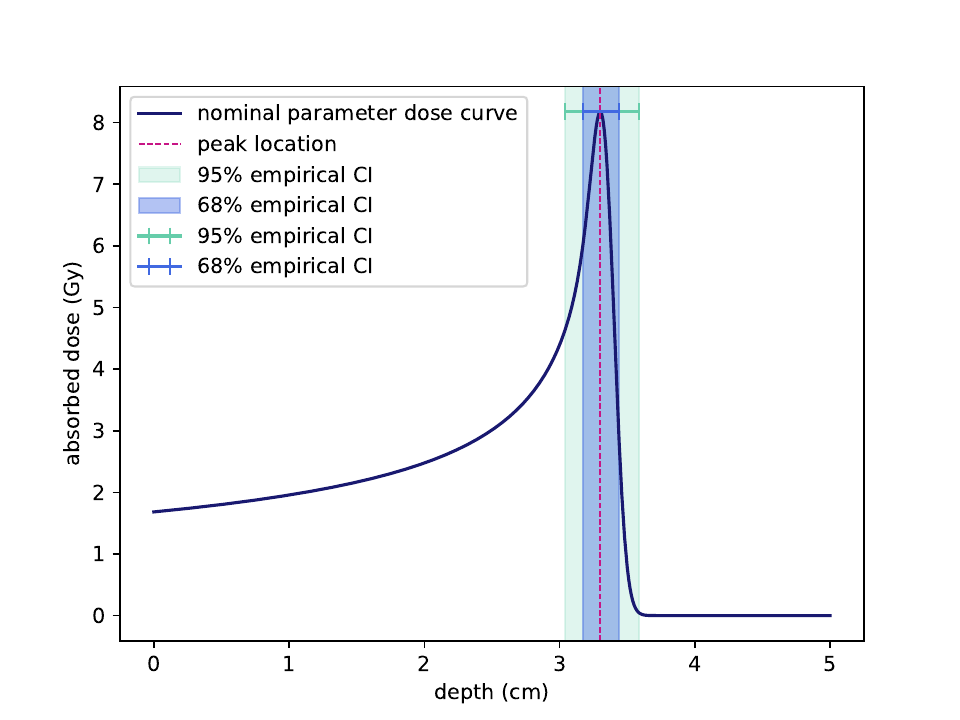}
    \includegraphics[width=0.3\linewidth]{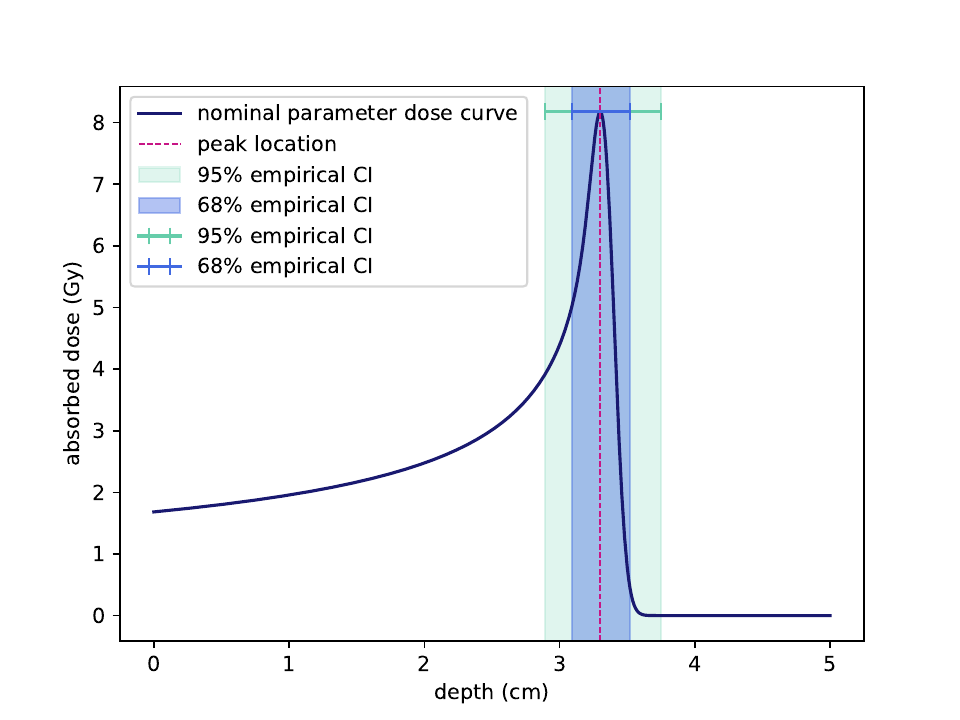}
    \caption{Confidence intervals for the variation in peak depth
      $z_{\text{peak}}$ when $\alpha$ and $p$ are normally distributed
      with means $\mu_{\alpha} = 0.00246, \mu_p = 1.75$ and standard
      deviations $\sigma_{\alpha} = 0.000128, \sigma_p = 0.0102$. The
      nominal dose curve $D(z;\mu_{\alpha},\mu_p)$, resulting from the assumed parameter values $\mu_{\alpha}$ and $\mu_p$, is plotted in dark blue.
      The corresponding 
      nominal peak depth of the curve $D(z;\mu_\alpha,\mu_p)$ is plotted as a dashed red line. The
      $68\%$ and $95\%$ confidence intervals for $z_{\text{peak}}$
      are shown as shaded regions, estimated using 25,000 independent
      random samples of $\alpha$ and $p$. On the left, only uncertainty in $\alpha$ has been included; in the centre only uncertainty in $p$, and on the right uncertainty in both $\alpha$ and $p$.}
    \label{fig:range_ci}
\end{figure}

In \autoref{fig:range_ci}, we show the estimated $68\%$ and $95\%$
confidence intervals for the peak position $z_{\text{peak}}$ of the
dose curve. These intervals are also derived using a Monte Carlo
approach with 25,000 independent random samples of $\alpha$ and $p$.

The results indicate that the confidence intervals for
$z_{\text{peak}}$ are comparable in magnitude when uncertainty is
included for either $\alpha$ or $p$ alone. However, when both
parameters are simultaneously uncertain, the confidence intervals for
$z_{\text{peak}}$ are considerably larger, consistent with the shifts
in $z_{\text{peak}}$ observed in
\autoref{fig:shifted_coefs}. Even relatively small uncertainties
in $\alpha$ and $p$ result in a $95\%$ confidence interval for
$z_{\text{peak}}$ of approximately $\pm 1$ cm, a level of uncertainty
significant for treatment planning. This highlights the need for
precise parameter estimation to ensure accurate dose delivery.

\newpage

\section{Treatment Planning}
\label{sec:optimisation}

To complete our study, we now examine how the model developed can be
applied to treatment planning in proton therapy.

The goal of treatment planning is to determine the optimal initial
beam angles, intensities, and energies such that the dose delivered to
a cancerous region is maximised while minimising the biological damage
to surrounding healthy tissues. This objective can be framed as a
control problem.

Efficient optimisation methods are crucial in Intensity-Modulated
Proton Therapy (IMPT), where practitioners often generate multiple
treatment plans along an approximate Pareto surface to balance
competing objectives and select the most suitable plan for the
patient. Furthermore, the optimisation of proton therapy treatment
plans involves a significantly larger number of decision parameters
compared to Intensity-Modulated Radiation Therapy (IMRT)
\cite{chen2010fast}.

\subsection{Dose Optimisation as a Constrained Least Squares Problem}

We consider the spatial domain divided into disjoint regions of
healthy and tumourous tissues, denoted $\Omega_H$ and $\Omega_T$,
respectively. For simplicity, we assume $\Omega_T = [z_{\text{prox}},
  z_{\text{dist}}]$. Given a target dose profile $T(x)$, the objective
is to construct an input beam $g: [E_{\text{min}}, E_{\text{max}}] \to
\mathbb{R}$ such that the resulting dose $D(z)$ closely approximates
$T(z)$.

To approach this problem, we require:
\begin{itemize}
    \item A discrete representation of the input beam $g$;
    \item A forward model $g \to D$ to predict the dose profile;
    \item A metric to quantify the difference between $D(z)$ and $T(z)$.
\end{itemize}

To represent the input beam, we assume it is a superposition of a
finite set of Gaussian-shaped basis beams, $\varphi_i$, $i = 1,
\ldots, N_y$, each centred at a principal energy $E_i$ with variance
$\sigma_i^2$. This formulation reduces the space of possible input
functions $g$ to a finite-dimensional vector space spanned by the
basis beams:
\begin{equation}
    g(\vec{y}, E) = \sum_{i=1}^{N_y} y_i \varphi_i(E),
\end{equation}
where $\vec{y} = (y_1, \ldots, y_{N_y})$ represents the weights or
intensities of each constituent beam. Each $\varphi_i$ approximates a
mono-energetic beam, and the choice of Gaussian-shaped beams ensures
practical feasibility given equipment constraints, as discussed in
\cite{markman2002beyond}. Alternative representations, such as
piecewise constant or linear approximations, may lead to input beams
that are difficult to realise in practice.

The forward model is provided by the one-dimensional analytical model
from \S\ref{sec:model}, which is linear with respect to the initial
beam energy $g$. This allows the total dose profile to be expressed as
a linear combination of precomputed dose profiles $D_i(z)$
corresponding to unit-intensity beams:
\begin{equation}
    D(\vec{y}, z) = \sum_{i=1}^{N_y} y_i D_i(z).
\end{equation}
By precomputing the dose profiles $D_i(z)$ for each basis beam
$\varphi_i$, the optimisation problem is reduced to finding the
optimal coefficients $\vec{y}$, which is computationally efficient.

To measure how well $D(\vec{y}, z)$ approximates $T(z)$, we define a
cost function that penalises deviations between the dose and the
target. Let $w(z)$ be a non-negative weighting function, then the cost
functional is
\begin{equation}
    l(\vec{y}) 
    :=
    \int_X w(z)\left(D(\vec{y}, z) - T(z)\right)^2 \, \d x.
\end{equation}
For practical implementation, we evaluate the dose at a finite set of
points $z_0, \ldots, z_{N_x}$ and approximate the integral using the
composite trapezoidal rule. This leads to the discrete cost functional
\begin{equation}
    L(\vec{y}) 
    := 
    \sum_{j=1}^{N_x} \frac{1}{2} w(z_j)\left(D(\vec{y}, z_j) - T(z_j)\right)^2 (z_j - z_{j-1}).
\end{equation}

\begin{definition}[Treatment planning optimisation problem]\label{def:opt}
  Given beams $\varphi_1, \ldots, \varphi_{N_y}$, target and weighting
  functions $T, w$, and an admissible set $Y$, find $\vec{y} \in Y$
  such that:
  \begin{equation}
    L(\vec{y}) \leq L(\vec{y}') \quad \forall \vec{y}' \in Y.
  \end{equation}
\end{definition}

\begin{remark}[Weighting function $w$]
  The weighting function $w$ provides flexibility in defining the cost
  function, allowing different priorities in the treatment plan. We
  illustrate this with the following examples:
  \begin{enumerate}
  \item 
    In the simplest case, where the goal is to deliver a specified
    dose to a target region $\Omega_T$ with no restrictions elsewhere,
    $w$ can be set as the indicator function of $\Omega_T$:
    \begin{equation}
      w = \mathbb{1}_{\Omega_T}.
    \end{equation}
  \item 
    When additional considerations, such as sparing an organ at risk
    (OAR) within $\Omega_O \subseteq \Omega_H$, are required, the
    weighting function can assign different priorities to regions. For
    example:
    \begin{equation}\label{eq:weighting}
      w(z) = 
      \begin{cases} 
        w_T & \text{for }  z \in \Omega_T, \\
        w_O & \text{for }  z \in \Omega_O, \\
        w_H & \text{otherwise}.
      \end{cases}
    \end{equation}
    Here, setting $w_O \gg w_T$ reflects a higher priority for sparing
    the OAR over achieving the target dose in the tumour.
  \end{enumerate}
\end{remark}

\begin{remark}[Admissible set $Y$]
  The admissible set $Y$ allows the inclusion of practical constraints
  on beam intensities. For instance, beams must have non-negative
  intensity, so $Y$ must satisfy:
  \begin{equation}
    Y \subseteq \{\vec{y} \in \mathbb{R}^{N_y} : y_i \geq 0 \,\, \forall i\}.
  \end{equation}
  Additionally, upper bounds on intensity may be enforced due to
  equipment limitations or safety constraints.
\end{remark}

\subsubsection{Selection of the Inflow Energy Profiles}

The success of the optimisation problem in \autoref{def:opt} depends
on appropriate choices for the beam profiles $\varphi_i$ and the
admissible set $Y$. Physically, it is desirable for the spread-out
Bragg peak (SOBP) to cover the tumour region $\Omega_T$. This requires
the ranges of the constituent beams $\varphi_i$ to lie within
$\Omega_T$.

The range of a proton beam, determined by the Bragg-Kleeman rule (see
Equation \eqref{eq:range_energy}), can be inverted to compute the
energy $E$ of protons with a given range $R$:
\begin{equation}
    E 
    = 
    \left(\frac{R}{\alpha}\right)^{1/p}.
\end{equation}
This relationship allows the selection of principal energies (the
centres of the Gaussian profiles for each beam) such that the ranges
satisfy:
\begin{equation}
    z_{\text{prox}} \leq \alpha E_i^p \leq z_{\text{dist}}.
\end{equation}

The choice of beam principal energies and widths significantly affects
the appearance of the resulting SOBP. As illustrated in
\autoref{fig:overlap_or_not}, restricting all beam ranges to lie
strictly within the tumour region can lead to oscillations at the
distal end of the SOBP. These oscillations persist even when the
boundary condition is resolved with a greater number of
beams. Allowing some beams with energies $E_i$ such that $R(E_i) >
z_{\text{dist}}$ reduces these oscillations but increases the dose
delivered to the surrounding healthy tissues.

\subsubsection{Example 1: Uniform Dose Delivery}

We let $\Omega_T = [3, 6]$ and $\Omega_H = X \backslash \Omega_T$. The
weighting function is chosen as described in \autoref{eq:weighting},
with $w_T = 1$ and $w_H = 0$. For this example, we set $N_y = 30$,
with beam energies $E_i$ selected such that their ranges are equally
spaced in $[z_{\text{prox}}, z_{\text{dist}} + \frac{1}{4}]$, and set
$\sigma_i^2 = 1$ for all $i$. The admissible set $Y$ is defined as:
\begin{equation}
    Y = \{\vec{y} \in \mathbb{R}^{N_y} : y_i \geq 0 \,\, \forall i\},
\end{equation}
to enforce non-negativity of beam intensities.

In practice, it is typical to deliver a homogeneous dose to the tumour
region \cite{grassberger2011variations}. As a first numerical
experiment, we define the target dose profile as:
\begin{equation}
    T(z) = 
    \begin{cases} 
        1 & \text{for } z \in \Omega_T, \\
        0 & \text{otherwise},
    \end{cases}
\end{equation}
and aim to find parameters $\vec{y}$ such that:
\begin{equation}\label{eq:l2_cost}
    \|w(x)^{1 / 2}\left(D(\vec{y}, z) - T(z)\right)\|_{L^2} \to \min.
\end{equation}

The optimisation problem is solved using the
Broyden–Fletcher–Goldfarb–Shanno (BFGS) algorithm. Results are shown
in \autoref{fig:unconstrained_dose}. The solution achieves highly
uniform coverage of the target region. Specifically, if $\vec{y}^*$ is
the optimal set of parameters, the relative error satisfies:
\begin{equation}
    \frac{L(\vec{y}^*)}{L((0,0,\ldots,0))} \approx 10^{-6},
\end{equation}
where:
\begin{equation}
    L((0,0,\ldots,0)) =
    \int_X w(z) T(z)^2 \, \mathrm{d} z.
\end{equation}

\begin{figure}[h!]
    \centering
    \includegraphics[width=\linewidth]{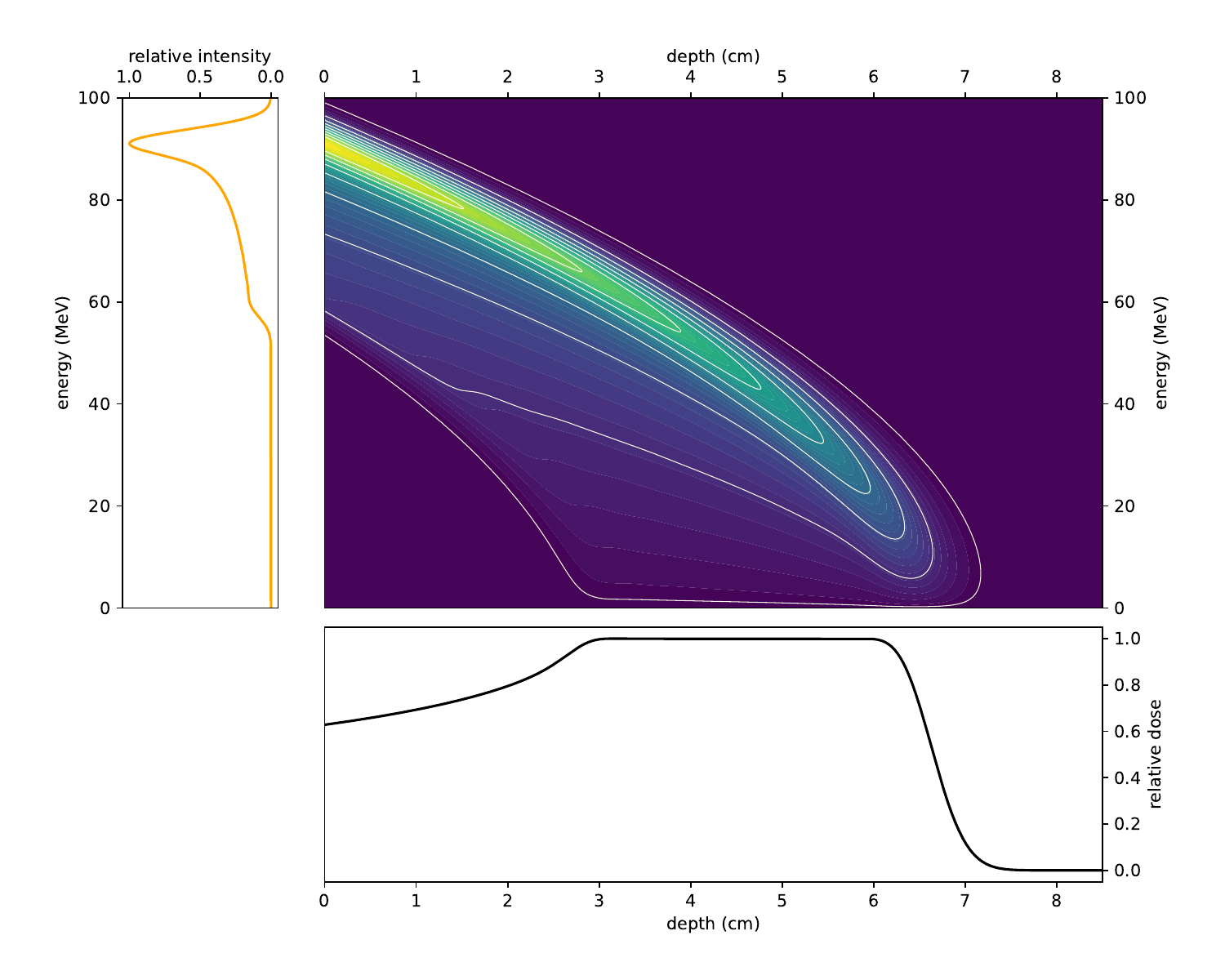}
    \caption{Visualisation of the input beam, fluence, and dose
      profile for Example 1. Left: the optimised input beam
      intensities across different energies. Middle: the fluence in
      depth-energy space, showing how the superposition of beams
      evolves through the medium. Right: the resulting dose profile as
      a function of depth, achieving a uniform dose within the target
      region $\Omega_T$.}
    \label{fig:sobp_prettyplot}
\end{figure}

\begin{figure}[h!]
    \centering
    \includegraphics[width=0.45\linewidth]{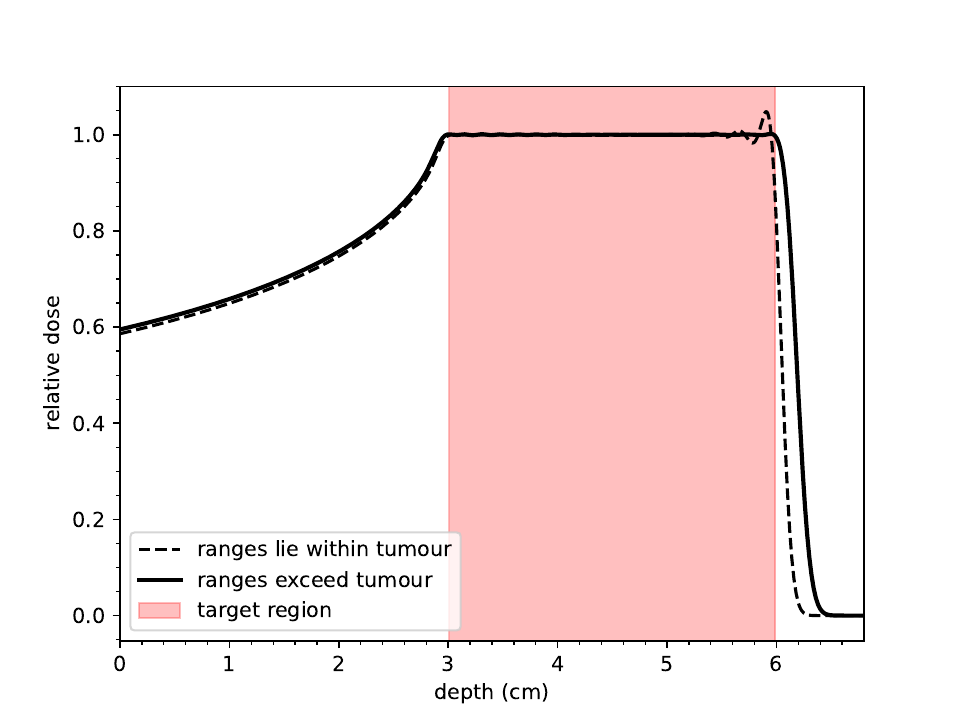}
    \includegraphics[width=0.45\linewidth]{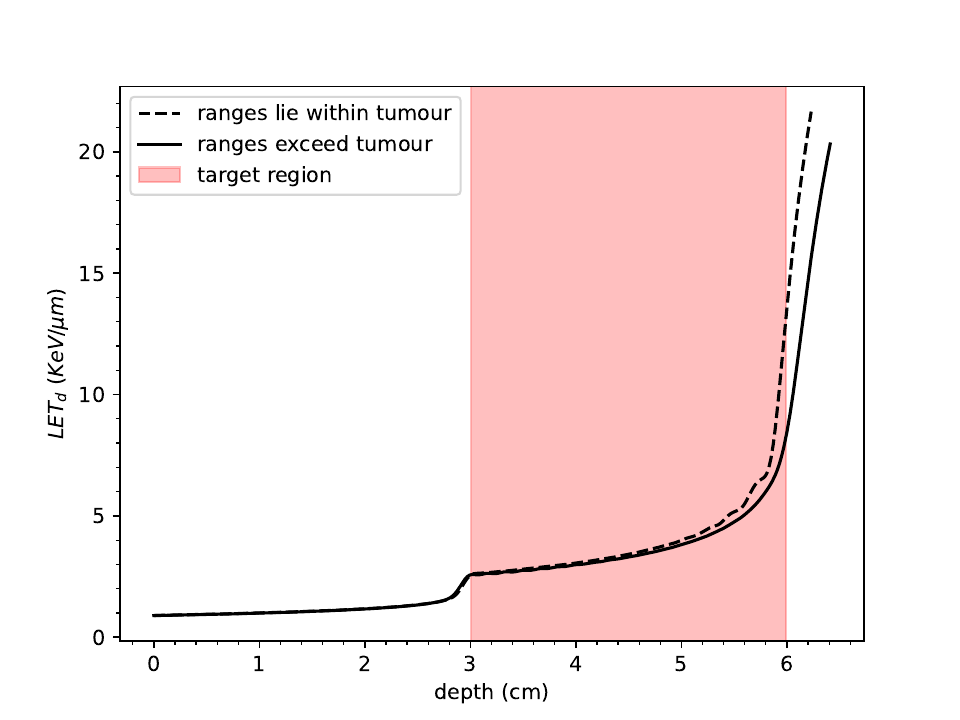}
    \caption{Spread-out Bragg peaks (left) and dose-averaged LET
      curves (right) resulting from the optimisation problem
      \autoref{def:opt}. Dashed lines: beam ranges are equally spaced
      between $z_{\text{prox}}$ and $z_{\text{dist}}$. Solid lines:
      beam ranges are equally spaced between $z_{\text{prox}}$ and
      $z_{\text{dist}} + 0.15$. Allowing slight extension of ranges
      into healthy tissue improves dose uniformity within the tumour.}
    \label{fig:overlap_or_not}
\end{figure}

\begin{figure}[h!]
    \centering
    \includegraphics[width=0.495\linewidth]{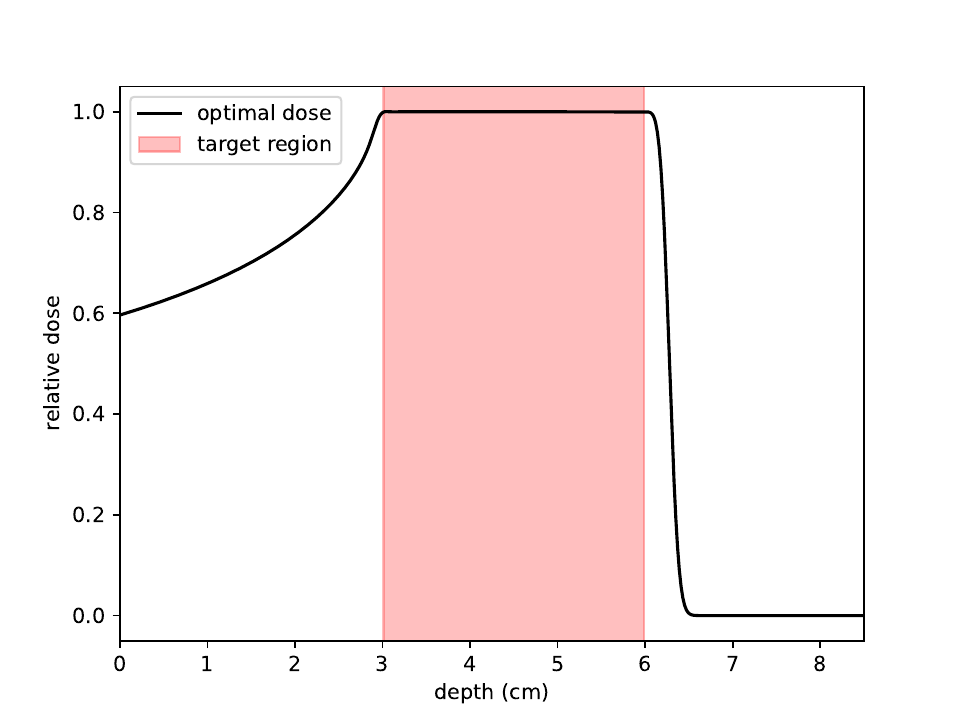}
    \includegraphics[width=0.495\linewidth]{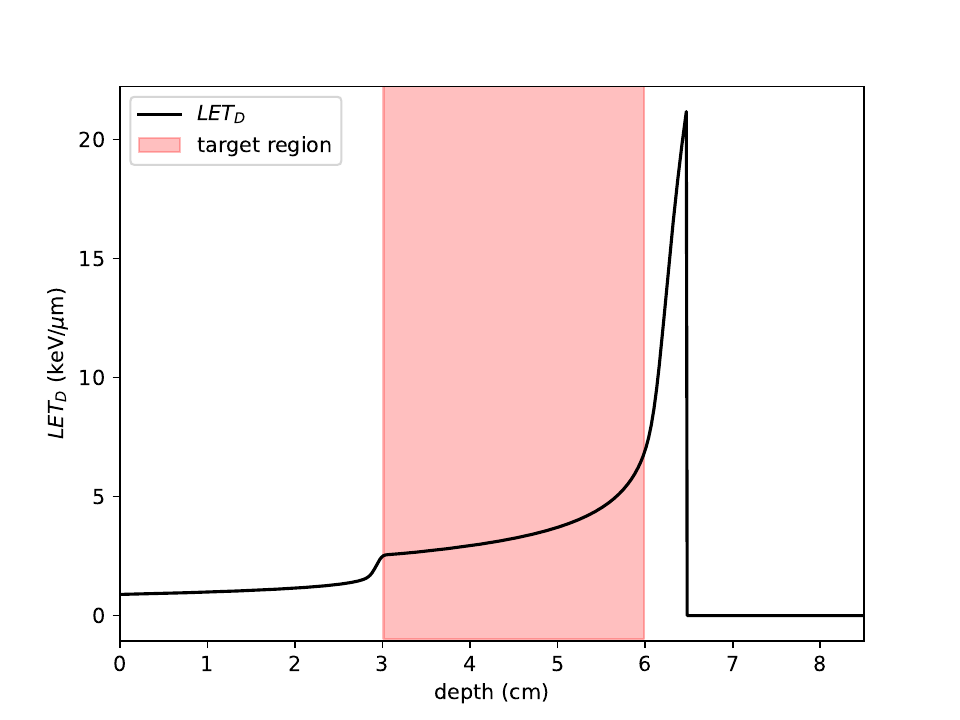}
    \caption{Left: dose profile resulting from the optimisation in
      Example 1. Right: corresponding dose-averaged LET. A total of 30
      energy levels are used, with energies chosen such that their
      ranges are equally spaced and cover the target region. Uniform
      dose delivery to the tumour is achieved.}
    \label{fig:unconstrained_dose}
\end{figure}

To provide a visual representation of the optimisation process and the
resulting dose distribution, \autoref{fig:sobp_prettyplot} illustrates
the input beam configuration, the fluence in depth-energy space, and
the final dose profile for Example 1. This figure parallels the
visualisation provided earlier for the pristine Bragg peak (see
\autoref{fig:prettyplot}), extending to the optimised spread-out Bragg
peak (SOBP) used in this treatment plan. This visualisation connects
the optimised input beam parameters to the resulting dose
distribution.

\subsubsection{Example 2: Organ at Risk (OAR)}

In this example, we retain the tumour region $\Omega_T = [3, 6]$ but
introduce an organ at risk (OAR) in $\Omega_O = [6, 8]$. The weighting
function $w$ is modified to penalise dose delivered to the OAR, and is
defined as:
\begin{equation}
    w(z) = 
    \begin{cases} 
        w_T = 1 & \text{for } z \in \Omega_T, \\
        w_O = 10 & \text{for } z \in \Omega_O, \\
        w_H = 0 & \text{otherwise}.
    \end{cases}
\end{equation}
As in Example 1, we set $N_y = 30$ and select beam energies such that
their ranges are equally spaced in $[z_{\text{prox}}, z_{\text{dist}}
  + \frac{1}{4}]$.

\begin{figure}[h!]
    \centering
    \includegraphics[width=0.495\linewidth]{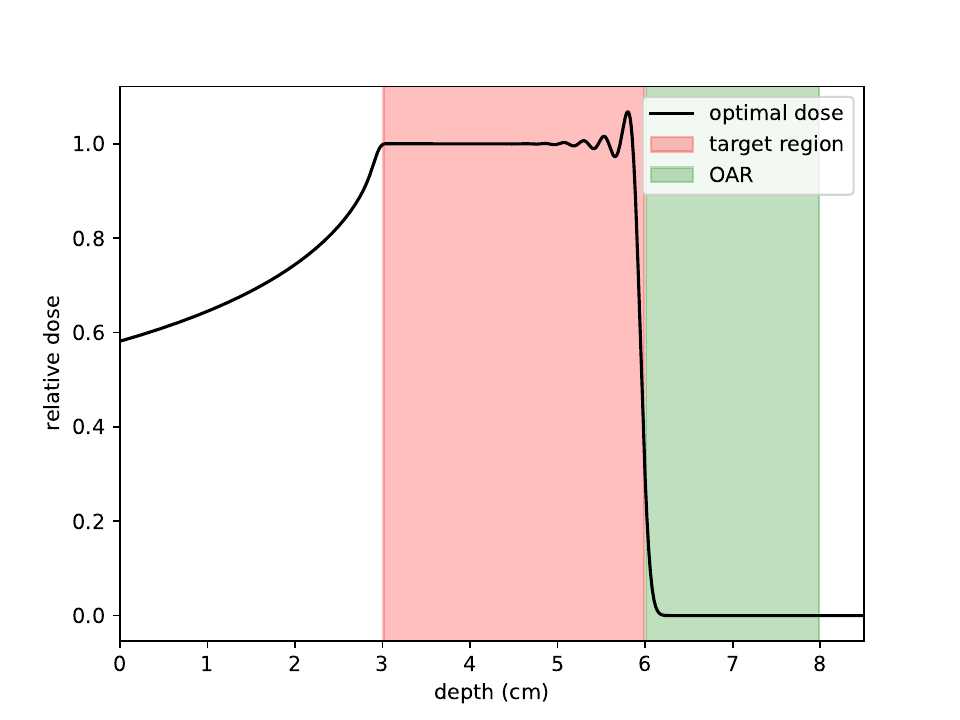}
    \includegraphics[width=0.495\linewidth]{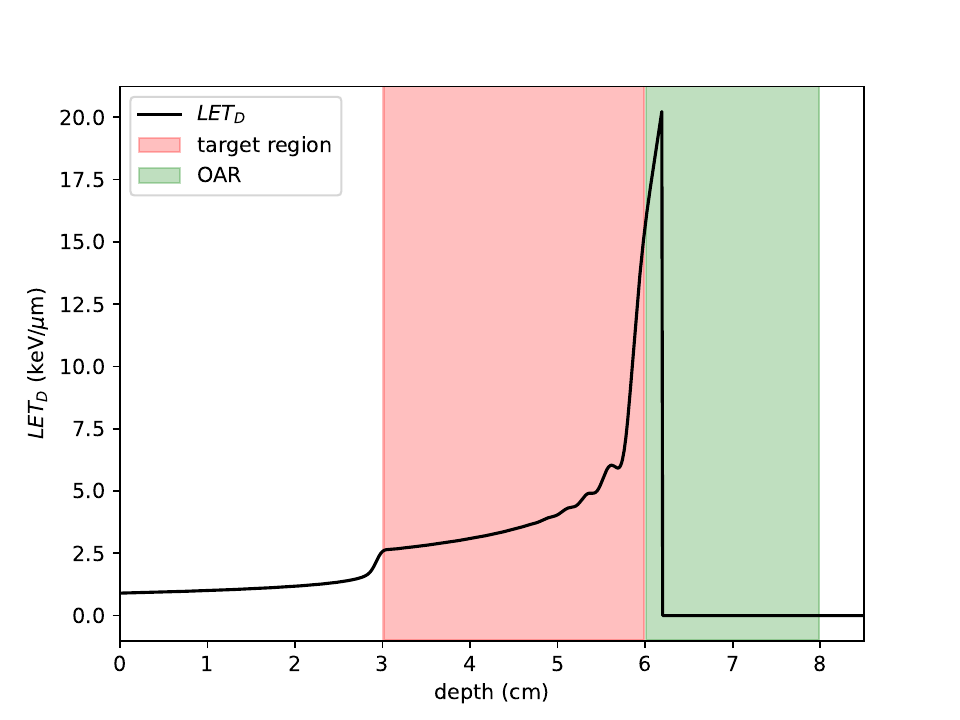}
    \caption{Left: dose profile resulting from the optimisation in
      Example 2. Right: corresponding dose-averaged LET. Penalising
      dose in the OAR (shaded green) reduces dose penetration into
      healthy tissue but introduces slight under-dosing and
      oscillation in the tumour's distal region.}
    \label{fig:oar}
\end{figure}

The results of this optimisation are shown in \autoref{fig:oar}. The
competing objectives of delivering sufficient dose to the tumour while
sparing the OAR result in less uniform dose coverage within the tumour
region. Compared to \autoref{fig:unconstrained_dose}, there is a
notable reduction in LET near the distal edge of the tumour,
accompanied by a significant decrease in dose delivered to the
OAR. However, this optimisation introduces oscillations and slight
under-dosing in the tumour’s distal region, which may be clinically
relevant depending on the treatment context.

\begin{figure}[h!]
    \centering
    \includegraphics[width=0.495\linewidth]{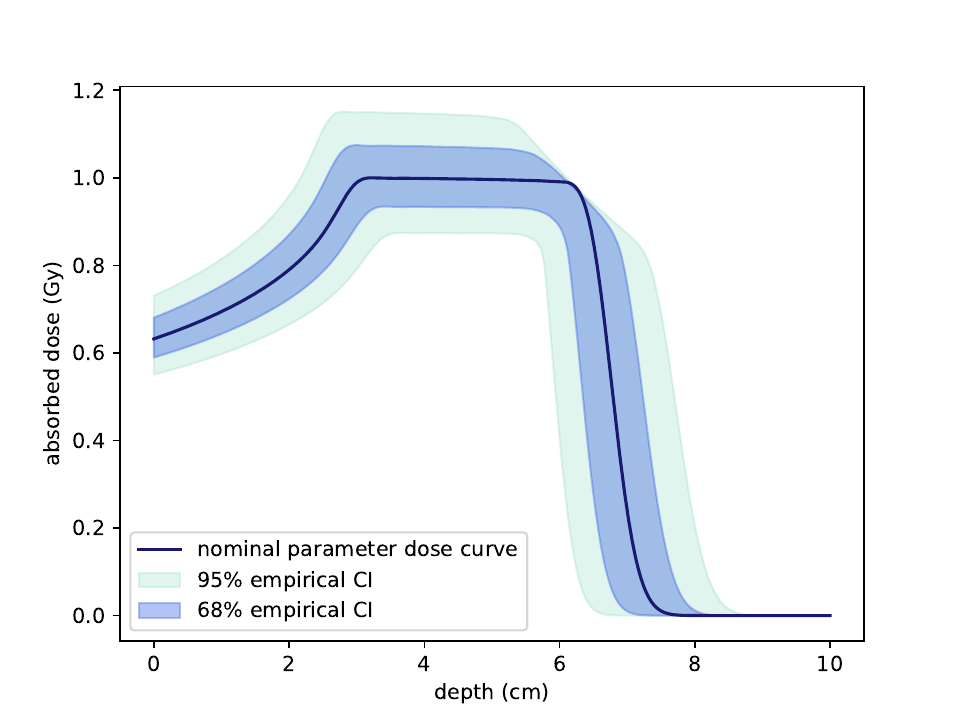}
    \includegraphics[width=0.495\linewidth]{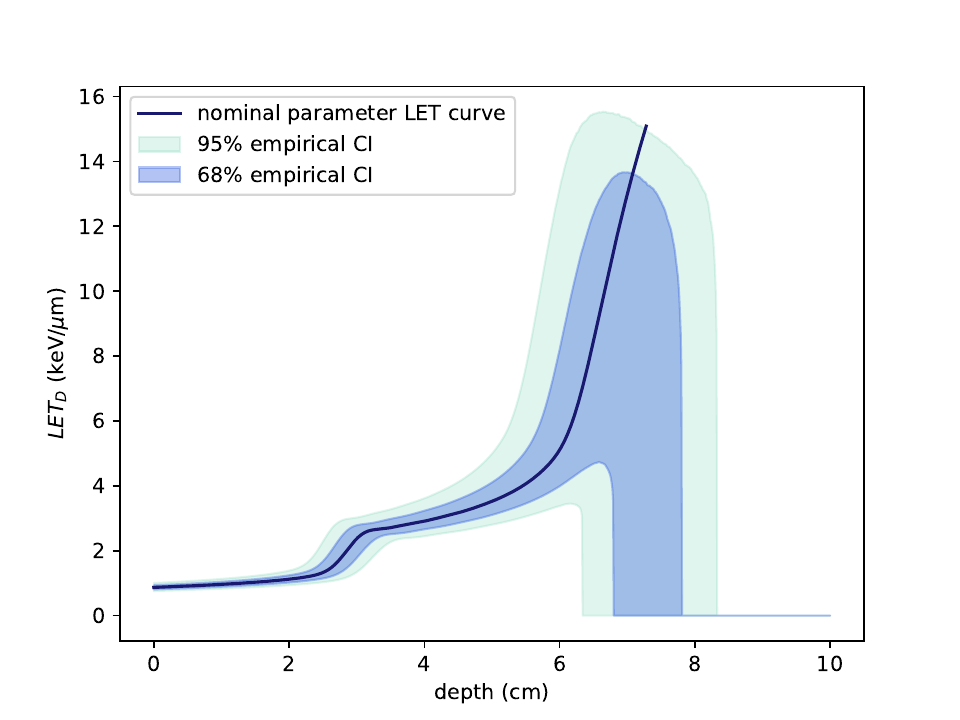}
    \caption{Confidence intervals for a spread-out Bragg peak and the
      corresponding dose-averaged LET. The confidence intervals are
      obtained as in \autoref{sec:uncertainties}, by introducing
      uncertainty in the stopping power parameters $\alpha$ and $p$,
      and estimating the empirical confidence intervals from 25,000
      independent samples using Monte Carlo methods. The dark blue
      curves represent the nominal dose and LET profiles, calculated
      using the assumed parameter values $\mu_{\alpha}$ and $\mu_p$.}
    \label{fig:sobp_ci}
\end{figure}

To further understand the implications of model uncertainty, we
examine the effects of parameter variability on the spread-out Bragg
peak (SOBP) for a scenario that includes an OAR. Figure
\ref{fig:sobp_ci} shows the Monte Carlo-estimated confidence intervals
for both the SOBP and the corresponding dose-averaged LET. The
methodology follows that described in \autoref{sec:uncertainties},
where uncertainty in the stopping power parameters $\alpha$ and $p$ is
introduced, and empirical confidence intervals are computed from
25,000 independent samples.

The results highlight significant uncertainty in the maximum dose
attained within the SOBP, as well as in the falloff region beyond the
tumour. Similarly, there is substantial uncertainty in LET,
particularly near the depth where the LET curve becomes sharply
peaked. These uncertainties are clinically important to consider in
treatment planning, as they heavily affect the balance between tumour
coverage and OAR sparing. When an OAR is located directly behind the
tumour, such uncertainties can compound the challenge of achieving an
optimal treatment plan.

\subsection{Optimisation Based on Biological Metrics}

An alternative approach to treatment planning involves prescribing the
fraction of surviving cells as a function of space, aligning the
optimisation process more closely with biological outcomes. For
instance, one could aim to kill 90\% of cells within the tumour,
resulting in an objective function of the form:
\begin{equation}
    L_{\mathcal{SF}}(\vec{y}) 
    :=
    \sum_{j=1}^{N_x}\frac{1}{2} \left(w(z)(\mathcal{SF}(\vec{y}, z) - T_{\mathcal{SF}}(z))^2\right)(z_j - z_{j-1}),
\end{equation}
where $\mathcal{SF}(\vec{y}, x)$ denotes the survival fraction and
$T_{\mathcal{SF}}(x)$ represents the target survival fraction.

While this approach is biologically motivated, it presents practical
challenges. Due to the exponential relationship between dose and
survival fraction, the optimisation problem becomes more
computationally expensive and can sometimes yield counterintuitive
results. For instance, regions receiving excessive dose may not incur
a significant penalty in the objective function, as the survival
fraction in those regions is already near zero.

To address these limitations, a biologically weighted dose metric is
often preferred. By incorporating biological weighting factors into
the dose, the optimisation problem becomes linear in terms of the
control parameters, significantly reducing computational
cost. Moreover, this approach avoids the spurious results associated
with excessive dose regions, providing a more robust framework for
treatment planning while retaining a biologically informed
perspective.

\subsubsection{Example 3: Uniform LET-Weighted Dose}

In this example, we optimise for a uniform biological dose within the
target region $\Omega_T$, defined as:
\begin{equation}
    T_{BD}(z) := 
    \begin{cases} 
        1 & \text{for } z \in \Omega_T, \\
        0 & \text{otherwise}.
    \end{cases}
\end{equation}

\begin{figure}[h!]
    \centering
    \includegraphics[width=0.495\linewidth]{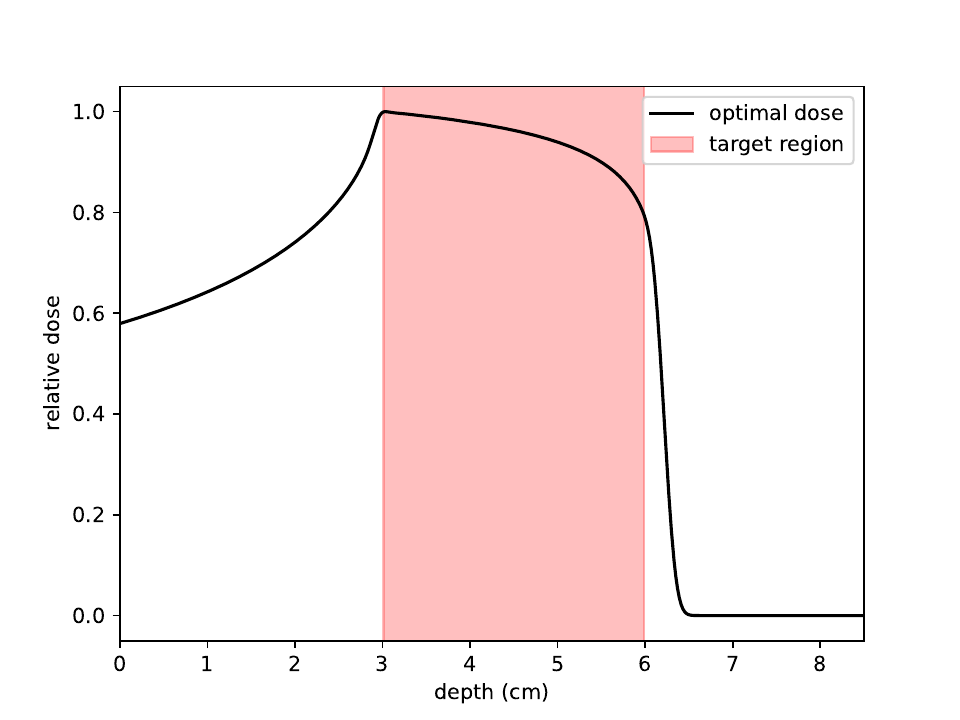}
     \includegraphics[width=0.495\linewidth]{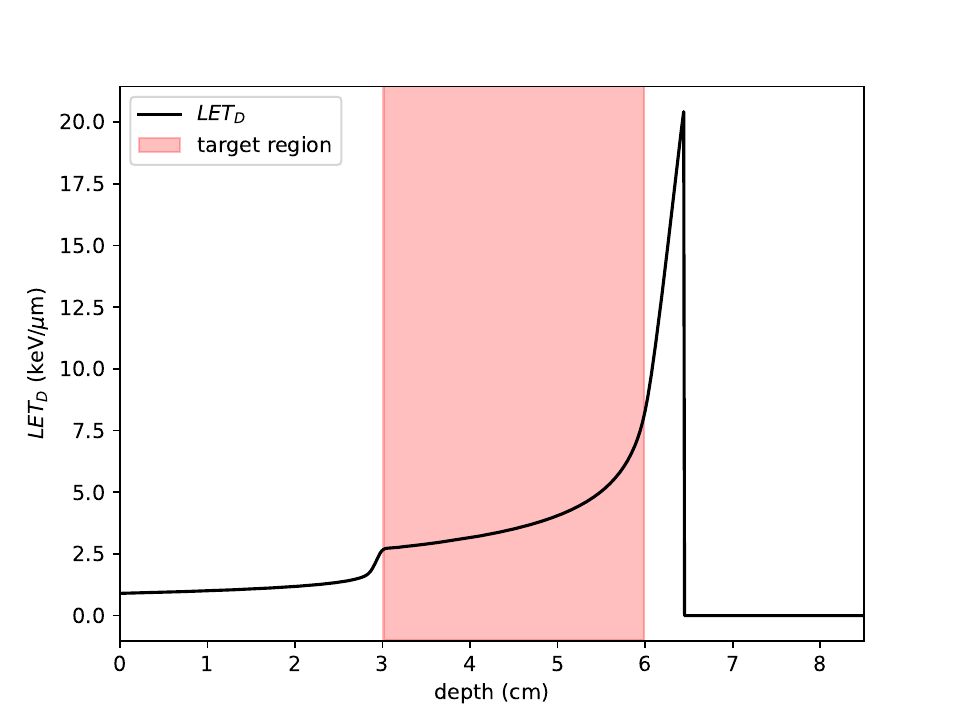}
     \caption{Left: dose profile resulting from the optimisation in
       Example 3. Right: corresponding dose-averaged LET. A total of 30
       energy levels are used, with energies chosen such that their
       ranges are equally spaced and cover the target region.}
    \label{fig:let_based_optimisation}
\end{figure}

The input beams are chosen as in Examples 1 and 2, with the weight
function $w$ set to $\mathbb{1}_{\Omega_T}$. The results of this
optimisation are shown in
\autoref{fig:let_based_optimisation}. Optimising for biological dose
introduces a trade-off: some dose conformity is sacrificed at the
distal part of the tumour to account for the higher LET values that
occur there. Consequently, the dose delivered is significantly lower
than that obtained by considering absorbed dose alone, reflecting the
heightened biological impact of protons near the Bragg peak.

\begin{figure}[h!]
   \centering
   \includegraphics[width=0.495\linewidth]{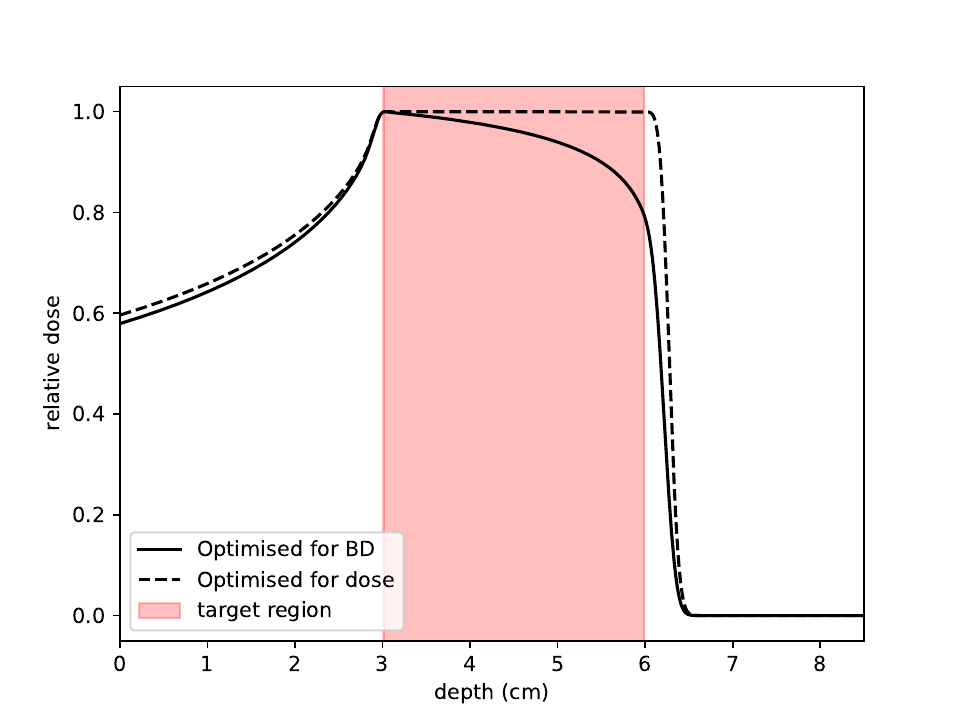}
   \includegraphics[width=0.495\linewidth]{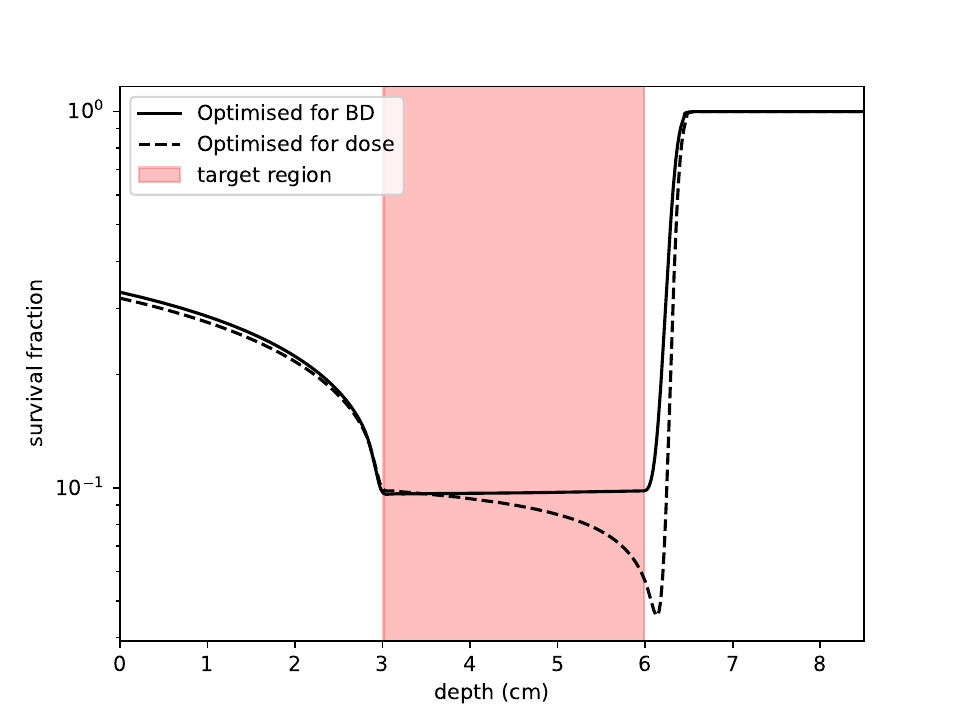}
   \caption{Comparison of relative dose (left) and survival fraction
     profiles (right) resulting from optimisation for absorbed dose
     (dashed line) and biological dose (solid line).  The survival
     fraction was computed using the linear-quadratic model, with the
     $\alpha$ parameter accounting for LET, for AG01522 cells using
     parameters from \cite{chaudhary2014relative}. Optimising for
     biological dose results in a tapered dose profile at the distal
     end of the target region, delivering significantly less dose to
     healthy tissue while maintaining an almost constant biological
     effect within the target.}
    \label{fig:different_sfs}
\end{figure}

Interestingly, the largest biological effect, computed using the
linear-quadratic model, is observed outside the target region when LET
is taken into account, as shown in \autoref{fig:different_sfs}. This
arises because the LET-weighted optimisation naturally prioritises
regions of higher biological effectiveness, even if they fall outside
the prescribed dose boundaries. A comparison of the survival fraction
profiles resulting from absorbed dose and biological dose optimisation
demonstrates this effect. The tapering of the dose profile towards the
distal end of the target region ensures significantly less dose is
delivered to healthy tissue, while maintaining a near-uniform
biological effect within the tumour.

It is instructive to compare these results with those in Figure 4 of
\cite{giovannini2016variable}, where a similar shape is observed in
RBE-weighted dose curves. The survival fraction profiles in
\autoref{fig:different_sfs} exhibit a comparable trend, underscoring
the alignment between LET-weighted dose optimisation and RBE-based
approaches in clinical practice.

\section{Conclusion}

In this work, we have developed a computationally efficient framework
for evaluating key metrics in PBT, including dose delivery, LET and
biologically informed metrics such as RBE and cell survival
fraction. Leveraging a simple analytical model, we achieve results
that show good agreement with those from computationally intensive
Monte Carlo particle simulations, while significantly reducing
computational cost. This makes the framework particularly well-suited
for rapid evaluations in treatment planning.

The speed and simplicity of the approach enable the exploration of
optimisation strategies with respect to challenging objectives, such
as LET-weighted dose or survival fraction. Such objectives, while
important for improving treatment outcomes, would require significant
computational resources if approached using Monte Carlo
simulations. Our framework allows for the efficient evaluation of
these biologically informed metrics, providing a practical tool for
exploring their potential integration into treatment planning
workflows as well as for investigating the impact of model
uncertainty, particularly in scenarios involving OARs.

By presenting these ideas in a mathematically rigorous but
approachable way, we believe this work also serves as an accessible
introduction for mathematicians interested in contributing to the
field of PBT. The integration of physical, biological and
computational principles offers a clear pathway for mathematical
researchers to engage with and address real-world challenges in cancer
therapy.

This work demonstrates the value of computationally fast models in
bridging the gap between theoretical modelling and practical
application in PBT. We believe it provides a foundation for future
investigations which account for other interaction mechanisms, Coulomb and nuclear, into biologically informed treatment planning, enabling
the rapid assessment of new metrics and strategies that may otherwise
be computationally prohibitive, supporting the broader goal of
delivering better patient outcomes in personalised cancer therapies.

\section*{Data availability statement}

The codebase used to generate the figures in this work is available at \url{https://doi.org/10.5281/zenodo.14179258}.

\section*{Acknowledgements}

The research was conducted by a working group sponsored
by the Radioprotection theme of the Institute for Mathematical
Innovation at the University of Bath. AP and TP were supported by the EPSRC
programme grant EP/W026899/1. BA and TP also received support from the
Leverhulme Trust grant RPG-2021-238 and TP the EPSRC grant
EP/X030067/1. VC is supported by a scholarship through the Statistical Applied Mathematics at Bath (SAMBa) EPSRC Centre for
Doctoral Training under the project
EP/S022945/1. DH is supported through an NPL iCASE scholarship. All this support is gratefully acknowledged.

\printbibliography

\end{document}